\documentclass[usegraphicx]{mn2e}
\usepackage{times}
\newlength{\colwidth}
\title[Radio imaging of the SXDF - I.]{Radio imaging of the Subaru/XMM-Newton
Deep Field - I. The 100$\mu$Jy catalogue, optical identifications, and the
nature of the faint radio source population}
\author[C.\ Simpson et al.]{Chris Simpson$^{1,2}$\thanks{E-mail:
cjs@astro.livjm.ac.uk}, Alejo Mart\'{\i}nez-Sansigre$^3$, Steve
  Rawlings$^3$, Rob Ivison$^{4,5}$, \newauthor
Masayuki Akiyama$^6$, Kazuhiro Sekiguchi$^6$, Tadafumi Takata$^7$,
Yoshihiro Ueda$^8$, \newauthor
and Mike Watson$^9$\\
$^1$Astrophysics Research Institute, Liverpool John Moores University,
Twelve Quays House, Egerton Wharf, Birkenhead CH41 1LD\\
$^2$Department of Physics, University of Durham, South Road, Durham
DH1~3LE\\
$^3$Department of Physics, University of Oxford, Denys Wilkinson Building,
Keble Road, Oxford OX1 3RH\\
$^4$UK Astronomy Technology Centre, Royal Observatory, Blackford Hill,
Edinburgh EH9~3HJ\\
$^5$Institute for Astronomy, University of Edinburgh, Royal Observatory,
Blackford Hill, Edinburgh EH9~3HJ\\
$^6$Subaru Telescope, National Astronomical Observatory of Japan, 650
N.~A`oh\={o}k\={u} Place, Hilo, HI 96720, USA\\
$^7$National Astronomical Observatory of Japan, Mitaka, Tokyo
181-8588, Japan\\
$^8$Department of Astronomy, Kyoto University, Kyoto 606-8502, Japan\\
$^9$Department of Physics and Astronomy, University of Leicester,
Leicester LE1 7RH}
\begin{document}

\date{Version of \today}

\pagerange{\pageref{firstpage}--\pageref{lastpage}} \pubyear{2006}

\maketitle

\label{firstpage}

\begin{abstract}
We describe deep radio imaging at 1.4\,GHz of the 1.3 square degree
Subaru/\textit{XMM-Newton\/} Deep Field (SXDF), made with the Very
Large Array in B and C configurations. We present a radio map of the
entire field, and a catalogue of 505 sources covering 0.8 square
degrees to a peak flux density limit of 100\,$\umu$Jy. Robust optical
identifications are provided for 90\,per cent of the sources, and
suggested IDs are presented for all but 14 (of which 7 are optically
blank, and 7 are close to bright contaminating objects). We show that
the optical properties of the radio sources do not change with flux
density, suggesting that AGNs continue to contribute significantly at
faint flux densities. We test this assertion by cross-correlating our
radio catalogue with the X-ray source catalogue and conclude that
radio-quiet AGNs become a significant population at flux densities
below 300\,$\umu$Jy, and may dominate the population responsible for
the flattening of the radio source counts if a significant fraction of
them are Compton-thick.
\end{abstract}

\begin{keywords}
catalogues --- galaxies: active --- radio continuum: galaxies --- surveys
\end{keywords}

\section{Introduction}

Surveys of extragalactic radio sources find two broad populations of
object: active galactic nuclei (AGNs), and star-forming galaxies. The
two processes which power these objects (accretion onto supermassive
black holes, and star formation) are believed to be intimately linked
by a mechanism or mechanisms collectively known as `AGN-driven
feedback' (e.g., Croton et al.\ 2006; Bower et al.\ 2006) which
produces the observed tight correlation between the masses of the
central black hole and the stellar bulge (e.g., Ferrarese \& Merritt
2000; Gebhardt et al.\ 2000). Radio surveys therefore provide the
opportunity to study the cosmic evolution of both these important
processes.

Spectroscopically-complete samples of radio sources exist at
equivalent 1.4-GHz flux densities of $\sim100$\,mJy (e.g., Willott et
al.\ 2002), and these samples are dominated by objects with powerful
radio-emitting jets which carry a kinetic power comparable to their
optical/X-ray photoionizing luminosity (e.g., Rawlings \& Saunders
1991). These objects are commonly referred to as `radio-loud' AGNs,
and have radio luminosities and morphologies which place them above
the Fanaroff \& Riley (1974) break ($L_{\rm178MHz} \sim 2 \times
10^{25}\rm\,W\,Hz^{-1}\,sr^{-1}$) and they are classified as FR
Class~II objects. At increasingly fainter flux densities, the observed
source counts follow a power-law to $\sim1$\,mJy (e.g., Windhorst et
al.\ 1993) and the spectroscopic completeness of samples selected at
these flux density limits decreases, in part because an increasing
fraction of the objects have spectra which show little sign of optical
nuclear activity (e.g., Waddington et al.\ 2001), apparently because
their supermassive black holes are accreting at a much lower
rate. While these are still genuine radio-loud AGNs due to their ratio
of optical and radio energy outputs, the radio luminosities place them
below the Fanaroff--Riley break as FR\,I sources.

The cosmic evolution of these two classes of radio-loud AGNs appears
to differ.  While it has been known for many years that the population
of the most powerful (FR\,II) extragalactic radio sources undergoes
very strong positive evolution (e.g., Dunlop \& Peacock 1990), the
less powerful (FR\,I) sources evolve less strongly, possibly with a
constant comoving space density (e.g., Clewley \& Jarvis 2004). By
modelling the evolution of the radio luminosity functions for these
sources, it can be shown that their contribution to the total radio
source counts should continue as a power law to even fainter flux
densities (e.g., Jarvis \& Rawlings 2004). However, below 1\,mJy the
observed counts flatten (when normalized to Euclidean), and this is
attributed to the appearance of a new population of radio sources
which do not contribute significantly at higher flux densities, but
which start to dominate at $S_{1.4} \la 300\,\umu$Jy.  This new
population is usually believed to be star-forming galaxies (Condon
1984; Windhorst et al.\ 1985), and the faint source counts have been
successfully modelled on the basis of this interpretation (e.g.,
Seymour et al.\ 2004). Recently, however, Jarvis \& Rawlings (2004)
have suggested that the source counts may flatten due to the
`radio-quiet' AGN population, by which we mean supermassive black
holes with a high accretion rate but a low ratio of radio to
optical/X-ray luminosity. For quasars, where the non-stellar optical
radiation is seen directly, the distinction between radio-loud and
radio-quiet can also be made on the basis of the ratio of radio to
optical fluxes (e.g., Kellermann et al.\ 1989).  Since radio-quiet
AGNs are known to be far more prevalent than their radio-loud
counterparts, this population could indeed be important.

The fact that we still do not know the nature of the population(s)
which dominate at flux densities $\la 300\,\umu$Jy can be attributed
to two factors. First, most of the effort to follow up radio source
samples spectroscopically to high completeness has been devoted to
samples selected at either bright ($\ga 100$\,mJy; e.g., Willott et
al.\ 2002) or very faint flux densities ($\la 50\,\umu$Jy) from
multi-wavelength studies of deep (and hence small-area) fields (e.g.,
Richards et al.\ 1998). As a result, there has been very little study
of the sources in the range $100\,\umu{\rm Jy} \la S_{1.4} \la
500\,\umu\rm Jy$, where the new population(s) begin to appear. Even
when studies have focused on this flux density range (e.g., Hopkins et
al.\ 2003), the optical counterparts span such a large range in
apparent magnitude that it is difficult to obtain IDs and subsequent
photometry and spectroscopy of all members of a sample. Results have
therefore been equivocal and even contradictory.  Windhorst et al.\
(1985) identified such sources with faint blue galaxies whose colours
suggested that they were undergoing significant star formation.
Spectroscopy by Benn et al.\ (1993) supported this interpretation, but
later work by Gruppioni, Mignoli \& Zamorani (1999) indicated that
early-type galaxies were the dominant population. They attributed this
difference to their deeper limit for optical identifications.

The key to solving this issue is therefore to increase the
identification fraction of faint radio surveys as close to 100\,per
cent as possible by combining deep multi-colour imaging and moderate
resolution (a few arcseconds, so as not to resolve out galactic-scale
starbursts and hence bias a sample in favour of active galaxies) radio
mapping over a degree-size field, so that there will be a
statistically useful number of such objects. The
Subaru/\textit{XMM-Newton\/} Deep Field (SXDF; Sekiguchi et al.\ 2001,
2006) is ideal for this task, as it is home to extremely deep optical
and near-infrared imaging (Furusawa et al.\ 2006; Lawrence et al.\
2006). Deep (50--100\,ksec) X-ray exposures with \textit{XMM-Newton\/}
(Ueda et al.\ 2006) have also been made in the SXDF, enabling the
active galactic nucleus (AGN) content of the radio source population
to be studied. We have therefore undertaken deep radio observations of
this field to enable the $\sim100\,\umu$Jy radio source population to
be identified and studied with greater completeness than has
previously been possible.

The format of this paper is as follows. In Section~2 we describe the
radio observations and reduction methods, and in Section~3 we explain
how we have determined and corrected for incompleteness in the
catalogue. In Section~4 we describe how optical identifications were
made, and report on our analysis of the faint radio source
population. In Section~5 we summarize our main results. Radio maps of
sources which are either significantly extended or have an uncertain
optical identification are presented in an Appendix. All magnitudes
quoted in this paper are calibrated on the AB scale.

\section{Observations and reduction}

\begin{figure*}
\resizebox{\hsize}{!}{\includegraphics{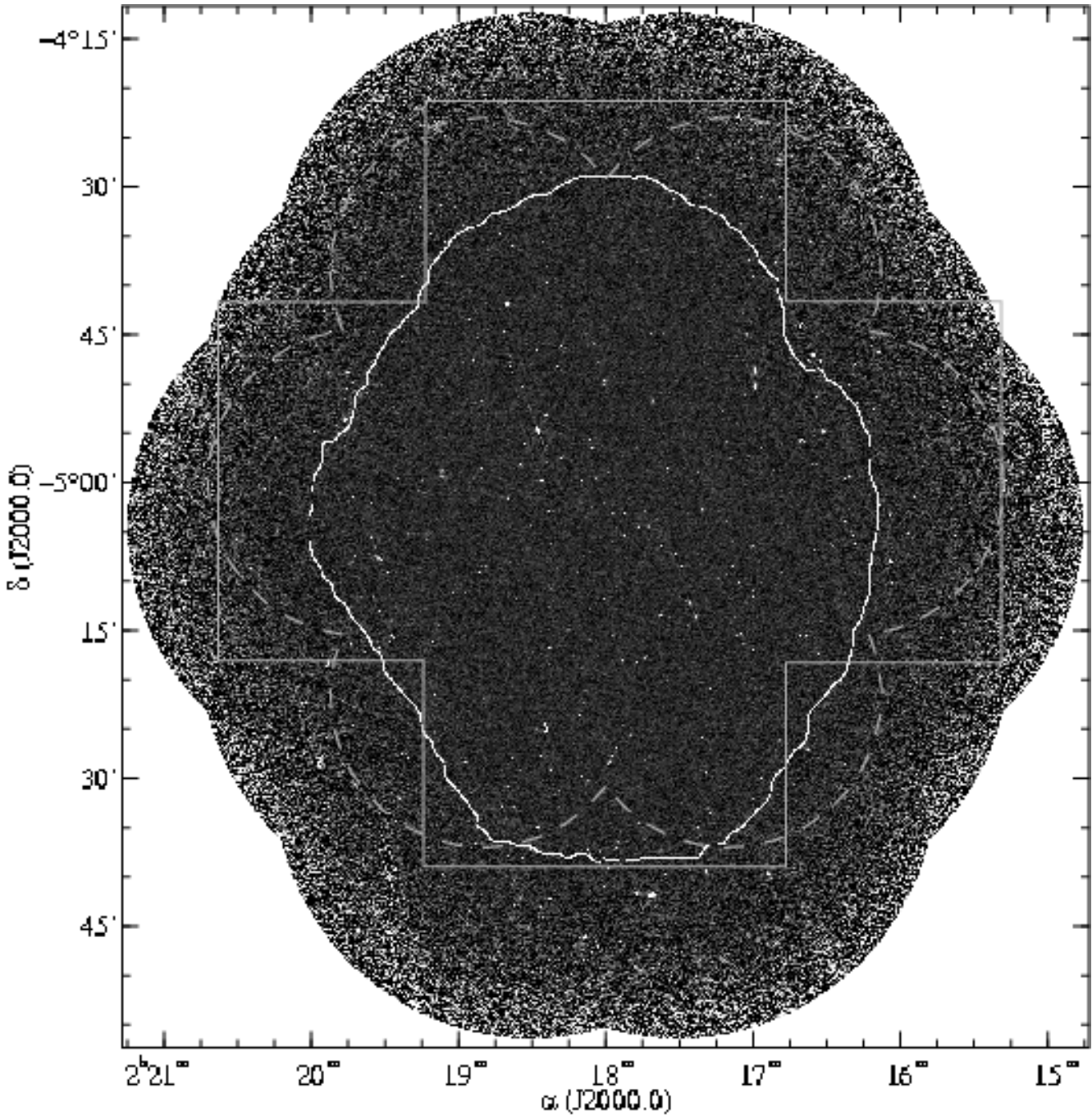}}
\caption[]{Final radio mosaic of the SXDF. The solid gray line shows
the region covered by the ultra-deep Suprime-Cam images (Furusawa et
al.\ 2006), while the dashed gray line shows the area observed with
the seven \textit{XMM-Newton\/} pointings (Ueda et al.\ 2006). The
solid white line indicates the region where the rms noise is less than
20\,$\umu$Jy\,beam$^{-1}$, from which the 100\,$\umu$Jy catalogue
described here has been constructed. Some mapping artefacts are still
evident.\label{fig:vlamap}}
\end{figure*}

The SXDF was observed with the National Radio Astronomy Observatory's
Very Large Array (VLA) in B-array using the 14 overlapping pointings
arranged in an hexagonal pattern listed in
Table~\ref{tab:vlacentres}. Three test observations of pointings 1, 4
and 6 were taken on 2001 May 17, and the rest of the data were
obtained in 13 runs, each lasting 4.5 hours, between 2002 August 10
and September 9. Each of these runs consisted of 5 snapshots of about
45-minutes duration separated by phase and bandpass calibration
observations of VLA calibrators 0217+017 and 0241$-$082 and a single
visit to the primary calibrator 3C~48 (0137+331). All data were taken
in the `4 IF' spectral line mode with central frequencies set at
1.3649 and 1.4351\,GHz, each with seven useable channels of width
3.125\,MHz per IF. The correlator delivered both `LL' and `RR'
correlated signals from the left (`L') and right ('R') circular
polarizations from each antenna, and a 5-second averaging time was
used. The maximum loss of amplitude (for objects 15 arcmin away from
the field centres) caused by the finite bandwidth and averaging time
is calculated to be 0.9\% and 0.02\%, respectively, and can be
considered negligible.

\begin{table}
\caption[]{Locations of the 14 field centres for the VLA
observations.\label{tab:vlacentres}}
\centering
\begin{tabular}{cccc}
Pointing & RA & dec & rms noise \\
& \multicolumn{2}{c}{(J2000.0)} & ($\umu$Jy) \\
\hline
1  & 02 18 30 & $-$05 04 20 & 20 \\
2  & 02 18 30 & $-$05 30 20 & 20 \\
3  & 02 17 30 & $-$05 04 20 & 21 \\
4  & 02 17 00 & $-$05 17 20 & 20 \\
5  & 02 19 00 & $-$04 51 20 & 21 \\
6  & 02 18 00 & $-$04 51 20 & 21 \\
7  & 02 19 30 & $-$05 04 20 & 20 \\
8  & 02 18 00 & $-$05 17 20 & 21 \\
9  & 02 17 00 & $-$04 51 20 & 21 \\
10 & 02 18 30 & $-$04 38 20 & 20 \\
11 & 02 17 30 & $-$04 38 20 & 22 \\
12 & 02 16 30 & $-$05 04 20 & 22 \\
13 & 02 17 30 & $-$05 30 20 & 20 \\
14 & 02 19 00 & $-$05 17 20 & 27 \\
\hline
\end{tabular}
\end{table}

All 14 pointings were re-observed in C-array on 2003 January 15 to
provide additional information on larger angular scales. Apart from
the array configuration, the set-up for these observations was
identical to that for the B-array observations and total integration
times for each pointing were around 5 minutes.

The summer 2002 B-array observations were severely affected by narrow-band
interference signals and much manual flagging of these datasets were
required.  After such editing, and standard phase and flux density
calibrations, data in each array and pointing positions were processed
using the $\cal AIPS$ task \textsc{imagr}. The CLEAN algorithm within
\textsc{imagr} worked on a continuous area of radius 0.52 degrees alongside
smaller fields centred on known NVSS sources (implemented using the
$\cal AIPS$ task \textsc{setfc}). The central region of each pointing
was divided into 13 facets which were eventually combined with the
$\cal AIPS$ task \textsc{flatn}. Maps made from the B-array data were
self-calibrated for phase (using the $\cal AIPS$ task \textsc{calib})
and then re-mapped with \textsc{imagr}: the clean component model of
the sky was also used to self-calibrate the C-array data before the B-
and C-array UV data were combined using the $\cal AIPS$ task
\textsc{dbcon}. Multiple cycles of running \textsc{imagr} and then
applying self-calibration techniques, initially on phase and finally
on phase and amplitude, were needed to improve the dynamic range of
each image. Various antennas with time-dependent problems were also
identified during this process and corrupted UV data were excised. A
consistent rms noise $\sim 20$--$22\,\umu$Jy was achieved in 13 of the
14 separately processed pointings, with an anomalously high noise level
$\sim 27 \rm \umu Jy$ in the pointing centred at 021900$-$051720; the
dynamic range of this pointing could not be improved despite extensive
effort. Finally, the 14 pointings were combined with \textsc{flatn} to
provide a final map in which the primary beam correction was
consistently applied.  This final image still has mapping artefacts
around the brightest sources, some of which might conceivably be
removable by more careful, manually intensive applications of the
CLEAN algorithm.  The synthesized beam varies slightly across the
image but has a roughly elliptical shape characterized by $\sim 5''
\times 4''$ at $\rm PA \approx 170^\circ$. We present the final
mosaic, and show its relationship to the areas covered by the optical
and X-ray data, in Fig.~\ref{fig:vlamap}.

\begin{figure}
\resizebox{\hsize}{!}{\includegraphics[angle=-90]{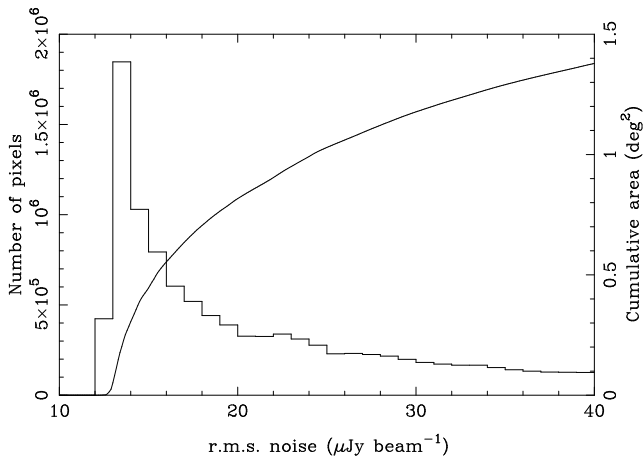}}
\caption[]{Distribution of pixel values (each pixel is 1.25$''$
square) in the noise map, as determined by SExtractor from
interpolation between the centres of the meshes. The cumulative area
covered by pixels with noise below a given value is shown by the
smooth curve.\label{fig:noise}}
\end{figure}

We calculate the rms noise as a function of position in our final
mosaic using the SExtractor software package (Bertin \& Arnouts
1996). A mesh size of $64\times64$ pixels, corresponding to
80\,arcseconds on a side, was used to compute the local background,
and the results are presented in Fig.~\ref{fig:noise}.

\section{Source counts}

For the remainder of this paper, we concentrate only on the region
where the rms noise per pixel, as determined from the Sextractor mesh,
is less than 20\,$\umu$Jy\,beam$^{-1}$. This region covers 0.808
square degrees (Figs~\ref{fig:vlamap} and \ref{fig:noise}) and is
well-matched to the area covered by the ultra-deep Suprime-Cam images
of this region (Furusawa et al.\ 2006).

\subsection{Source extraction}

The $\cal AIPS$ task \textsc{sad} was used to find and extract all the
sources with a peak flux density $\geq 100\,\umu$Jy, corresponding to
signal-to-noise ratios of between five and eight. A conservative cut was
chosen to minimize the appearance of spurious sources. The positions from
the \textsc{sad} catalogue were used to create the final SXDF radio
catalogue as described below.

\begin{figure}
\resizebox{\hsize}{!}{\includegraphics{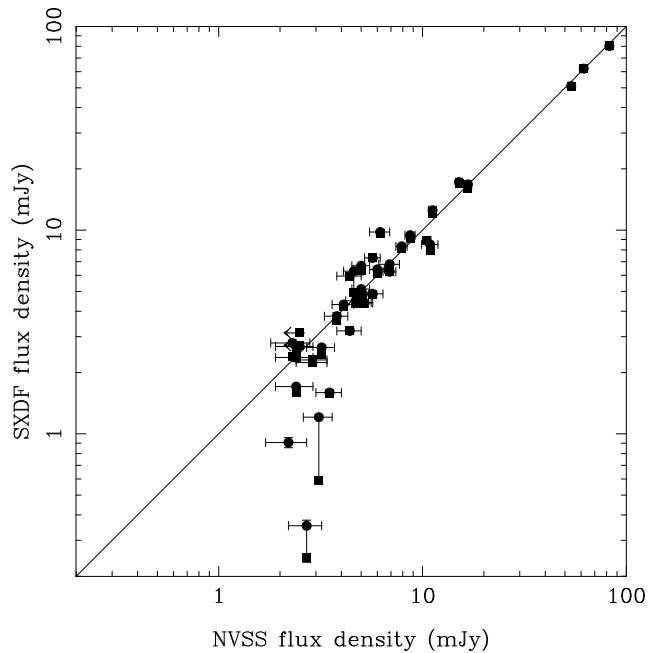}}
\caption[]{Comparison of our flux densities with those from the NVSS
catalogue (Condon et al.\ 1998). All but one of the NVSS sources can
be associated with a source in our catalogue, and we plot these as
squares. We have also investigated the effect of confusion caused by
the large beam of the NVSS, by including the flux from nearby sources
in our catalogue (weighted by a Gaussian with ${\rm FWHM} = 45''$),
and we plot these measurements as circles (the NVSS source not
uniquely identified with a source in our catalogue appears to be due
to the confused flux from a number of faint sources). The solid line
indicates an equality of flux densities.\label{fig:nvss}}
\end{figure}

Since \textsc{sad} works by fitting and then subtracting Gaussians to
the radio image, it is liable to two sources of error. First, it
systematically underestimates the flux of extended sources which
contain diffuse emission. Sources which were clearly extended were
dealt with by first defining masks to exclude them from the
\textsc{sad} catalogue, and then using the $\cal AIPS$ task
\textsc{tvstat} to measure their flux densities. It was possible to do
this on an object-by-object basis due to their small number (14);
examples include VLA~J021724$-$05218 and VLA~J021827$-$04546. To
investigate possible biases in the flux density measurements of the
brightest radio sources, we compared our measurements with those of
the NVSS (Condon et al.\ 1998), and we find no systematic effects
except for those NVSS sources at the survey flux limit
(Fig.~\ref{fig:nvss}).

Secondly, as discussed in detail by Condon (1997), Gaussian fitting
routines such as \textsc{sad} tend to systematically underestimate the
angular sizes and overestimate the peak flux densities when there is
noise present. We have attempted to produce an algorithm which derives
unbiased flux density estimates for all sources. First, every source
in the \textsc{sad} catalogue had its flux density measured in a
series of concentric circular apertures of diameter 7.5, 10, 12.5,
$\ldots$, 35\,arcseconds, as well as by the \textsc{imfit} task. The
flux densities in the 12.5 and 35-arcsecond apertures and the
\textsc{imfit} measurement were compared for each source, and if any
pair of these measurements disagreed by more than 2$\sigma$, the
source was flagged as possibly confused, and a flux density was
measured by hand using the \textsc{tvstat} task.

The following procedure was used to determined the 1.4-GHz flux
densities of the remaining sources. Each source in the \textsc{sad}
catalogue has a peak flux density $S_{\rm peak}$ and FWHM measurements
along the major and minor axes, $\theta_{\rm M}$ and $\theta_{\rm m}$,
respectively.  Uncertainties, $\sigma_{\rm M}$ and $\sigma_{\rm m}$
can be assigned to these measurements, following Condon (1997) and
Condon et al.\ (1998). These angular sizes are compared to the size of
the \textsc{clean} beam, $\theta_{\rm M}^* \times \theta_{\rm m}^*$.
A source was considered to be unresolved if
\[
\left[ {\rm max} \left( 0, \frac{\theta_{\rm M}-\theta_{\rm
M}^*}{\sigma_{\rm M}} \right) \right] ^2 +
\left[ {\rm max} \left( 0, \frac{\theta_{\rm m}-\theta_{\rm
m}^*}{\sigma_{\rm m}} \right) \right] ^2 \leq 2^2 \, ,
\]
in which case it was assigned a flux density of
\[
S_{\rm unres} = S_{\rm peak} \left(
\frac{\theta_{\rm M} \theta_{\rm m}}{\theta_{\rm M}^* \theta_{\rm
m}^*} \right) ^{1/2}
\]
(Condon 1997; Condon et al.\ 1998), with an appropriately calculated
uncertainty $\sigma_{\rm unres}$. Due to the tendency of \textsc{sad}
to overestimate flux densities, 17 sources in the original catalogue
were found to be unresolved but have flux densities $S <
100\,\umu$Jy. These sources were not included in the final catalogue.

\begin{figure}
\resizebox{\hsize}{!}{\includegraphics{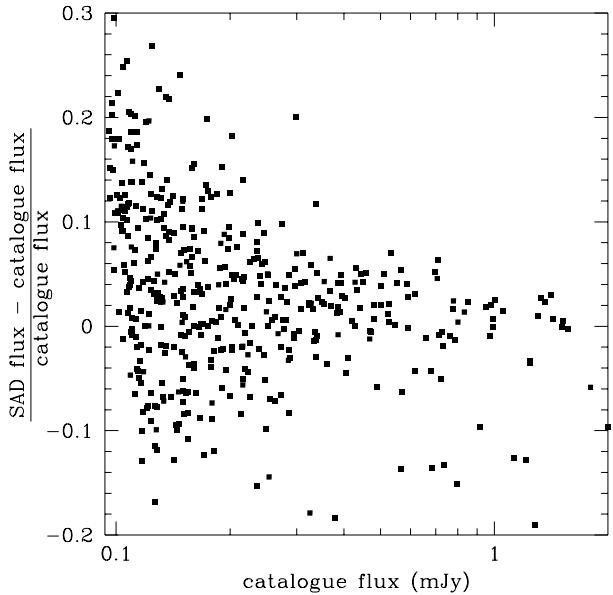}}
\caption[]{Relationship between the source flux densities as
determined by the $\cal AIPS$ task \textsc{sad} and the flux densities
used in the catalogue, demonstrating the bias in the task. Sources
significantly below the main locus of points are
extended.\label{fig:sadbias}}
\end{figure}

Additionally, an aperture flux density, $S_{\rm ap}$, was assigned to
each source using the concentric aperture measurements: working
outwards from the smallest aperture, when the measurements in two
consecutive annuli were both consistent (at the 1$\sigma$ level) with
zero, the flux density measured within the boundary between these
annuli was assigned. Sources with $S_{\rm ap} \geq S_{\rm unres} + 2
\sigma_{\rm unres}$ were inspected by eye and the aperture flux was
used in the catalogue unless the difference was due to confusion.
This method should ensure that sources with strong unresolved cores
and fainter extended emissions do not have their flux densities
systematically underestimated. The aperture flux was also used for all
sources which were not considered to be unresolved by the above
criterion. We demonstrate in Fig.~\ref{fig:sadbias} the systematic
difference between our catalogue fluxes and the biased \textsc{sad}
fluxes.

In a number of cases, widely-separated multiple components of a single
source were found by the detection algorithm. These instances were
identified by overlaying the radio contours on a true-colour optical
image of the field. In this manner, it became easy to distinguish
FR\,II double radio sources (where there was no optical counterpart at
the location of either radio-emitting component, but there was a
plausible ID -- usually a red galaxy -- between them) from instances
where distinct (but possibly physically related) sources of similar
radio flux density had a small angular separation. More detail on the
optical identification will be given in Section~4, but we note that a
method independent of the optical data, such as that suggested by
Bondi et al.\ (2003), would not work for our catalogue, as it would
both fail to identify large double sources, and also incorrectly
associate distinct sources having small angular separations. In total,
there are 505 unique sources in the final catalogue.

\subsection{Completeness}

Two issues need to be considered when interpreting the measured source
count distribution. First is the well-known effect of Eddington bias
(Eddington 1913), where random measurement errors result in more
objects being boosted into a given bin of flux density from the one
below than are removed from that bin by having a measured flux density
lower than the value (provided the source counts have the correct sign
of slope). This effect can be modelled since the observed source
counts, $N'(S)$, are a function of the true source counts, $N(S)$, and
the distribution of pixel values in the noise map. If $\sigma$ is the
rms noise of a pixel and $f(\sigma)\,d\sigma$ is the fraction of
pixels in the noise map with values between $\sigma$ and
$\sigma+d\sigma$, then
\begin{equation}
N'(S) = \int_0^{\infty} \int_{-\infty}^{\infty} \frac{1}{\sqrt{2\pi}}
{\rm e}^{-\zeta^2/2} N(S+\zeta\sigma) f(\sigma) \, {\rm d}\zeta \,
{\rm d}\sigma \, .
\label{eq:1}
\end{equation}
This equation cannot be inverted, but we can use a range of plausible
functional forms for $N(S)$ to examine the amount by which source counts
will be overestimated due to Eddington bias, i.e., $N'(S)/N(S)$.

\begin{figure}
\resizebox{\hsize}{!}{\includegraphics[angle=-90]{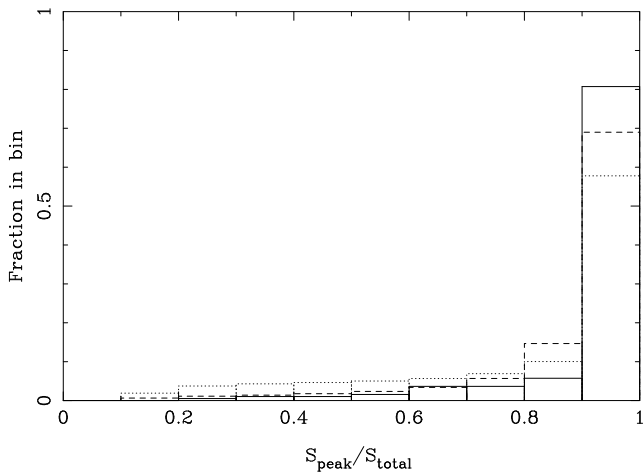}}
\caption[]{Histogram of peak-to-total flux ratio for the 192 sources
with $0.2\leq S_{\rm t}<1.0$\,mJy (solid line), compared with the
predicted distribution assuming the angular size distributions of Bondi
et al.\ (2003; dashed line) and Windhorst et al.\ (1990; dotted
line). The predictions have been normalized to account for the
incompleteness to very extended sources in our data.\label{fig:peaktot}}
\end{figure}

The second effect is the incompleteness to extended sources, since a
source only appears in our catalogue if its peak flux density is
greater than our threshold of 100\,$\umu$Jy\,beam$^{-1}$. To account
for this requires knowledge of the distribution of peak-to-total flux
ratios, $S_{\rm peak}/S_{\rm total}$, as a function of source flux
density. The form of this function is obviously dependent on the size
of the beam but, in principle, can be derived from the distribution of
source angular sizes. The shape of this distribution is
controversial. Windhorst, Mathis \& Neuschaefer (1990) claim that the
integral distribution of angular source sizes (i.e., the fraction of
sources larger than a given size $\psi$) has the form
\begin{equation}
h(\psi) = 2^{-(\psi/\psi_{\rm med})^{0.62}} \, ,
\end{equation}
where the median source size is given by
\begin{equation}
\psi_{\rm med} = 2.0'' (S_{1.4}/1{\rm\,mJy})^{0.3} \, ,
\end{equation}
while Bondi et al.\ (2003) claim that this overpredicts the number of
large ($\psi>4''$) sources by a factor of $\sim2$. While the Windhorst
et al.\ parametrization is based on a small number of sources, we
suspect that the different is due, at least in part, to different beam
sizes used by the authors. At sub-mJy flux densities, many radio
sources comprise both a relatively compact component and more
extended, diffuse emission and at high angular resolution, these
components will be separated and the ``angular size'' determined from
a fit to the compact component. At lower resolution, the entire source
is likely to be well-fit by a single component and so the measured
angular size will be larger. By convolving the distribution of source
sizes with our beam, we can predict the distribution of peak-to-total
flux ratios for sources with a given flux density, and we find that
the Bondi et al.\ form for $h(\psi)$ provides better agreement with our
observations (Fig.~\ref{fig:peaktot}).

\begin{table}
\caption[]{Source counts in the 100-$\umu$Jy catalogue.\label{tab:counts}}
\centering
\begin{tabular}{ccrcr}
$S$ & $\overline{S}$  & \multicolumn{1}{c}{$N$} & 
Completeness & $S^{2.5} dN/dS$ \\
(mJy) & (mJy) & & & ($\rm Jy^{1.5} sr^{-1}$)\\ \hline
0.100--0.126 & 0.112 &  98 & 0.775 & $2.67\pm0.27$ \\
0.126--0.158 & 0.141 &  90 & 0.963 & $2.77\pm0.29$ \\
0.158--0.200 & 0.178 &  63 & 0.982 & $2.68\pm0.34$ \\
0.200--0.251 & 0.224 &  50 & 0.989 & $2.99\pm0.42$ \\
0.251--0.398 & 0.316 &  74 & 0.994 & $3.69\pm0.43$ \\
0.398--0.631 & 0.502 &  40 & 0.999 & $3.97\pm0.63$ \\
0.631--1.000 & 0.795 &  31 & 1.002 & $6.12\pm1.10$ \\
1.000--3.162 & 1.809 &  33 & 1.000 & $8.96\pm1.56$ \\
3.162--10.00 & 5.801 &  20 & 1.000 & $31.18\pm6.97$ \\
\hline
\end{tabular}
\end{table}

\begin{figure*}
\resizebox{\hsize}{!}{\includegraphics[angle=-90]{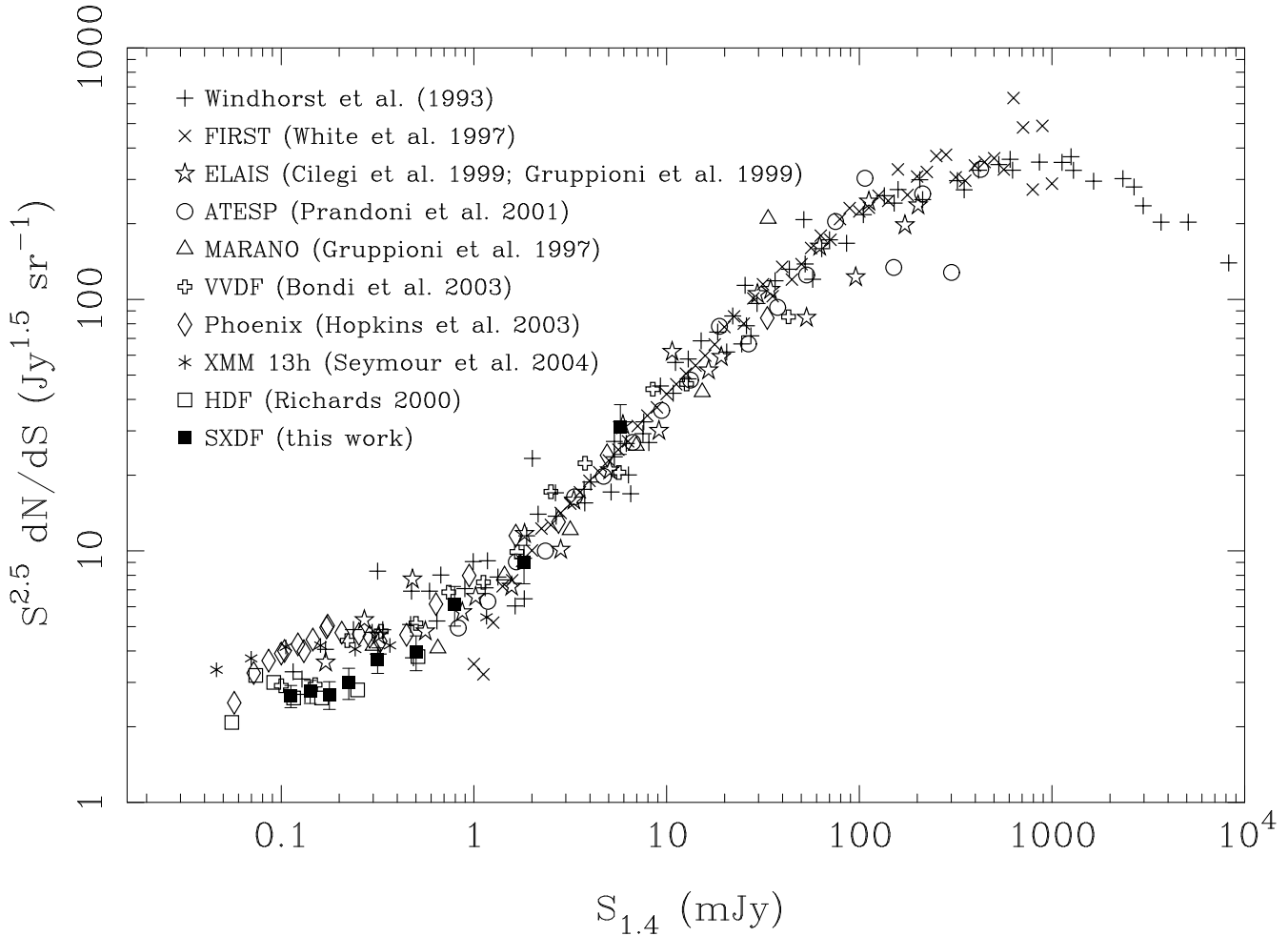}}
\caption[]{Differential source counts at 1.4\,GHz from this work (solid
symbols) and surveys in the literature: the compilation by Windhorst et
al. (1993); FIRST survey (White et al.\ 1997); ELAIS (Ciliegi et al.\ 1999;
Gruppioni et al.\ 1999); ATESP (Prandoni et al.\ 2001); MARANO (Gruppioni
et al.\ 1997); VLA-VIRMOS deep field (Bondi et al.\ 2003); Phoenix Deep
Field (Hopkins et al.\ 2003); \textit{XMM/ROSAT} 13--hour field (Seymour,
McHardy \& Gunn 2004); Hubble Deep Field North (Richards
2000).\label{fig:counts}}
\end{figure*}

Modifying Equation~\ref{eq:1} to include only sources where the
measured peak flux density exceeds our 100-$\umu$Jy limit, based on
the distribution of $S_{\rm peak}/S_{\rm total}$, allows us to
properly determine the incompleteness corrections which need to be
applied to our source counts. We adopt different forms for the true
source counts to determine how sensitive the correction factors are to
our assumed form for $N(S)$. We consider the sixth order polynomial of
Hopkins et al.\ (2003), the third order polynomial of Katgert, Oort \&
Windhorst (1998), and a form where the source counts are flat in a
Euclidian plot. The differences between these cases is negligible
(less than one per cent), and we use the corrections derived from the
Hopkins et al.\ parametrization. We also find that the correction
factors are insensitive to the choice of angular size distribution
and, while the use of the Windhorst et al.\ form does lead to an
overestimation of the correction factors, this is only true when the
factors are relatively small (at flux densities $\sim 1.5$ times the
detection threshold) and is not significant for a catalogue of our
size. In the bin $126 \leq S < 158\,\umu$Jy, for example, the
completeness is calculated to be 91\,per cent or 96\, per cent for the
Windhorst and Bondi models, respectively, and this difference only
exceeds the Poisson uncertainty for a large catalogue with more than
400 sources in this range (our catalogue has only 90). At higher flux
densities, the catalogue is essentially complete, while closer to the
threshold, the correction factor is determined by the fraction of
resolved sources which, for our beam size, is close to the median
source size (which is the same for both angular size
distributions). The completeness is higher if the functional form of
Bondi et al.\ scales with the median angular size, but even then the
difference is less than 10\,per cent and still comparable with the
Poissonian uncertainty. Deep A-array data (Ivison et al.\ 2006) have
been taken in the central region of the SXDF to identify submillimetre
sources detected in the SCUBA Half-Degree Extragalactic Survey
(SHADES; Mortier et al.\ 2005), and this will allow us to determine
more accurately the angular size distribution of sources at
$\sim100\,\umu$Jy levels, which we will do in a later paper. At the
present time, we determine correction factors to the observed source
counts from the Bondi et al.\ form for the angular size distribution,
assuming this does not change with flux density. The sources are
grouped in bins of flux density of width $\Delta \log S = 0.1$ at the
lowest flux densities, with an increasing bin size as the numbers
decrease towards higher fluxes. The centre of each bin was calculated
from equation 19 of Windhorst et al.\ (1984) where the slope of the
source counts was iteratively determined by interpolating between the
adjacent bins. The raw source counts and correction factors are given
in Table~\ref{tab:counts}, while the differential counts are presented
in Fig.~\ref{fig:counts}.

We note that there is almost a factor of two peak-to-peak variation in
the observed number counts at a flux density of $\sim200\,\umu$Jy
between the different surveys plotted in Fig.~\ref{fig:counts}.  The
SXDF points lie at the low extreme, but are consistent with those
measured from the HDF, VVDF, and Windhorst et al.\ (1990, 1993)
surveys. The Phoenix Deep Survey (Hopkins et al.\ 2003) has higher
number counts, despite covering approximately one square degree (the
\textit{XMM-Newton\/} 13-hour survey and ELAIS surveys also produce
much higher number counts, but cover significantly smaller areas).
Although the restoring beam used by Hopkins et al.\ is nearly four
times larger in area than ours (whereas the SXDF and VVDF use
restoring beams of similar size), the source density at these flux
limits is still too low for source confusion to be an important
effect. We can also rule out the possibility that we are missing
diffuse sources which are seen with a larger beam, since (i) the
Windhorst et al.\ surveys use an even larger beam, and (ii) we have
corrected for sources with peak flux densities below our limit, and so
have a good measurement of the total number of sources with
$S>100\,\umu$Jy. With $\sim100$ sources per flux density bin, the
statistical uncertainty is $\sim10$\%, while we can estimate the
effects of cosmic variance using the method of Somerville et al.\
(2004). Assuming that faint radio sources have the same correlation
length as mJy sources from the FIRST and NVSS surveys ($r_0 \approx
5$\,Mpc; Overzier et al.\ 2003) and sample the redshift interval
$z=1\pm0.5$, the r.m.s.\ uncertainty from cosmic variance is estimated
to be $\sim9$\% (equation~3 of Somerville et al.). A similar value is
obtained if a bias of $b\approx2$ (e.g., Peacock \& Dodds 1994) is
assumed for the radio sources. The extrema of the source counts
plotted in Fig.~\ref{fig:counts} may therefore represent $\pm2\sigma$
deviations from the mean number density.

\section{Optical identifications}

\subsection{Assignment of optical counterparts}

We have attempted optical identifications of the radio sources from the
ultra-deep \textit{BRi$'$z$'$\/} Suprime-Cam images of the SXDF. These are
described in detail by Furusawa et al.\ (2006). Seven sources lie outside
the region covered by these data, and for these sources we have used
shallower unpublished $R$ and $i'$ images from Suprime-Cam.

\begin{figure}
\resizebox{\hsize}{!}{\includegraphics{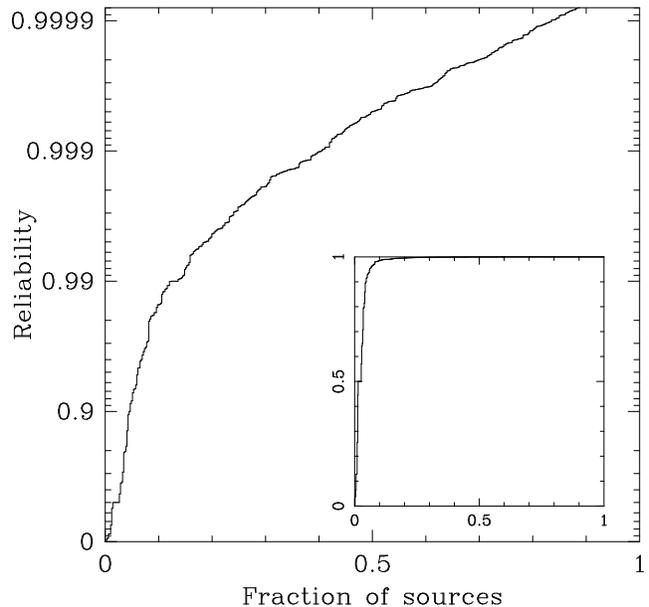}}
\caption[]{Histogram showing the fraction of sources with reliability
measurements (Sutherland \& Saunders 1992) above a given value. The
inset shows the same plot with a linear $y$-axis. Not included in this
figure are 7 sources where the optical identification is obscured by a
bright foreground object. The 15 sources whose radio morphologies made
the reliability measurement an inappropriate statistic have been
assigned values of 0.99, 0.9, and 0.5 if their identifications are
considered  \textit{reliable\/}, \textit{probable\/}, and
\textit{plausible\/}, respectively.\label{fig:lrplot}}
\end{figure}

We first confirm the astrometry by comparing the radio and optical
($R$-band) positions of those sources with unresolved (or
barely-resolved) morphologies in both bands. We select the 20
brightest such radio sources (all with peak signal-to-noise ratios
greater than 15) and find that the mean offsets in right ascension and
declination between the centroids in the two images are 0.03 and
0.24\,arcsec, respectively (with the optical IDs lying to the
southwest of the radio sources). Since these are both much smaller
than the size of the radio beam, we do not make any corrections to the
astrometry of either the optical or the radio images.

Likely identifications were then made for all sources by overlaying the
radio contours onto a true-colour \textit{BRz$'$\/} image of the field
around each radio source. A 2-arcminute field was used to ensure that all
possible double radio sources were identified. The relatively high
resolution of the radio imaging made most identifications unambiguous,
since there was typically only one source within an arcsecond of the peak
of the radio emission, even at extremely faint optical magnitudes. In
instances where more than one optical counterpart was visible, blue
unresolved sources (often identified with X-ray sources from the catalogue
of Ueda et al.\ 2006) or faint red resolved objects were favoured. For
cases where a source was composed of multiple radio components but no core
was visible, an identification was sought in the centre of the source.

Once an optical counterpart had been identified, it was matched to an
object in the Version~1 catalogue from the optical imaging. Usually, a
counterpart was found in the $R$-band catalogue, but occasionally the
object was blended and a different band had to be used. In some cases, the
proposed ID was blended in all the catalogues, in which case SExtractor was
re-run over the Suprime-Cam image with a more severe deblending threshold.

We investigate the reliability of our optical identifications using the
criterion of the likelihood ratio as defined by Wolstencraft et al.\
(1986), after de Ruiter, Arp \& Willis (1977), and developed by Sutherland
\& Saunders (1992) for cases where there may be more than one possible
identification for a given source. For each radio source $r$, likelihood
ratios $L_{or}$ are calculated for all optical sources $o$ within 5\,arcsec
of the radio position, using the formula
\begin{equation}
L_{or} = \frac{Q(<m_o)\exp(-r_{or}^2/2)}{2\pi \sigma_x \sigma_y N(<m_o)}
\, .
\end{equation}
Here, $Q(<m_o)$ is the fraction of radio sources whose optical
counterparts are brighter than that of the proposed optical ID (we
estimate this in an iterative fashion from our own catalogue);
$\sigma_x$ and $\sigma_y$ are the combined (radio plus optical)
1-sigma uncertainties in the positions along the major and minor axes
of the restored beam, respectively, calculated as $\sigma_x =
\sqrt{(0.45\theta_x/{\rm SNR})^2+0.29^2}$ (and similarly for
$\sigma_y$) using the formula of Reid et al.\ (1988) and adding in
quadrature the standard deviation measured from the radio--optical
offsets. $r_{or} = \sqrt{(\Delta x/\sigma_x)^2 + (\Delta
y/\sigma_y)^2}$ is the normalized separation between the radio and
optical positions, and $N(<m_o)$ is the surface density of objects
brighter than the proposed optical ID (determined from the Suprime-Cam
catalogues). We apply the offset measured between the radio and
optical astrometric frames before calculating the separation between
the radio and optical positions.  The analysis is performed in the
filter in which the initial identification was made (i.e., usually the
$R$-band).

\begin{figure}
\resizebox{\hsize}{!}{\includegraphics{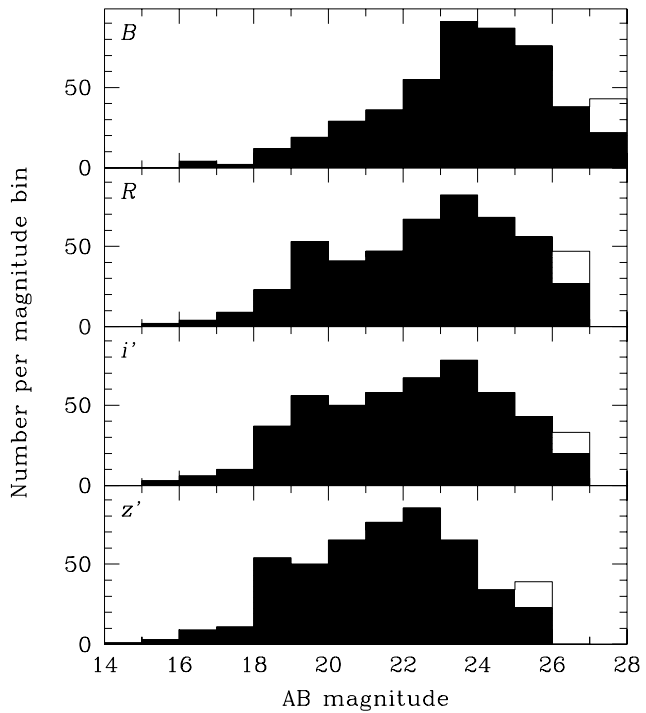}}
\caption[]{Magnitude distributions for the identifiable radio sources. The
open region in the faintest magnitude bin (the last bin where the
completeness of the optical catalogue is still high) indicates objects with
magnitudes fainter than this bin. In addition to the 7 sources whose
identification is hampered by the presence of bright foreground objects, a
further seven sources lie outside the multicolour SXDF imaging and only
appear in the $R$ and $i'$ histograms.\label{fig:maghists}}
\end{figure}

The probability than a particular optical object $i$ is the correct
identification for a radio source (the reliability) is then given by
\begin{equation}
{\cal R}_i = \frac{L_{ir}}{\displaystyle \sum_o L_{or} + (1-Q)}
\end{equation}
(Sutherland \& Saunders 1992), where $Q$ is the fraction of radio
sources with optical counterparts, which we estimate to be 95\,per
cent from our data (i.e., the fraction of sources with ${\cal R} \leq
0.5$, although the vast majority of the identifications are
insensitive to this value). Of our 505 sources, 7 cannot be identified
due to the presence of a bright foreground star or galaxy, while a
further 15 have extended radio morphologies for which the radio
position may not be indicative of the location of the optical
counterpart. In 13 of these cases, we assign optical identifications
based on our experience studying extragalactic radio sources and label
these identifications as \textit{reliable\/} (5), \textit{probable\/}
(1), or \textit{plausible\/} (6). One more (VLA~0016) has two possible
IDs, while two (VLA~0017 and VLA~0049) have \textit{certain\/} optical
counterparts. All these sources are presented in the Appendix. Formal
probabilities can be assigned to the proposed optical counterparts of
the remaining 483 radio sources by the above method. We assign as the
\textit{primary\/} optical counterpart the source with the largest
reliability, but in the catalogue we list all counterparts with ${\cal
R} > 0.05$. The catalogue is available on the world wide web and a
portion is shown in Table~\ref{tab:cat}.

Fig.~\ref{fig:lrplot} shows the cumulative distribution of reliability
measurements for all 498 sources not affected by foreground objects. A
total of 459 sources have probabilities of $>$90\,per cent of having
been correctly identified (plus a further 6 are \textit{reliably\/}
classified by eye, while 2 more have \textit{certain\/}
identifications) and, after the sorting the objects with
numerically-assigned reliabilities in order of decreasing probability,
there is a less than 50\,per cent chance of \textit{any\/} of the
first 436 sources being incorrectly identified. In the Appendix, we
present radio--optical overlays of the extended radio sources and
those where we have made an identification which has a low formal
probability.

\begin{figure}
\resizebox{\hsize}{!}{\includegraphics{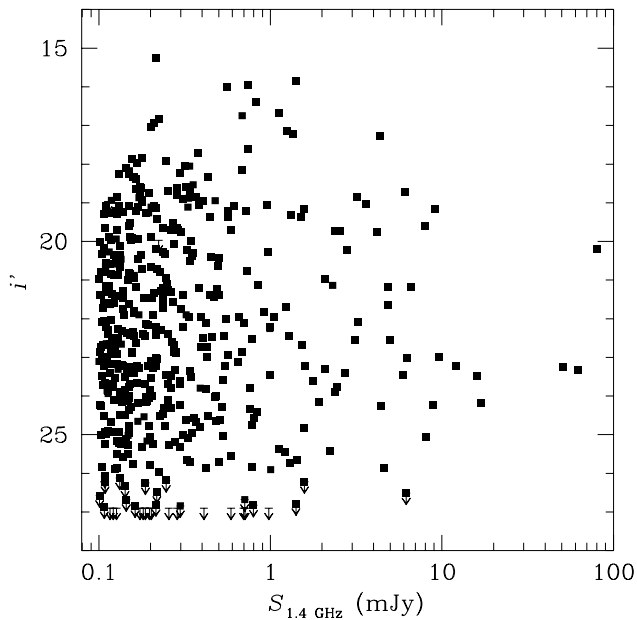}}
\caption[]{Optical ($i'$) magnitude vs radio flux density for all
sources not contaminated by bright foreground objects. Upper limits
indicate sources with no counterpart or whose identifications have
${\cal R} < 0.01$.\label{fig:optrad}}
\end{figure}

Our optical identification fraction of $\sim 90$\% (for sources with
${\cal R} \geq 0.9$ reduces by a factor of more than 2 the percentage
of `blank field' sources, compared to other surveys with shallower
optical limits (e.g., Sullivan et al.\ 2004; Ciliegi et al.\ 2005).
This can be naturally explained as being due to the greater depth of
our optical imaging. On the other hand, we also have a higher
identification fraction than Afonso et al.\ (2006) determined for
faint radio sources in the GOODS-S field using \textit{Hubble Space
Telescope\/} Advanced Camera for Surveys (ACS) imaging. Of the 37
radio sources with $S_{1.4} \ge 100\,\umu$Jy, they fail to find
optical identifications for 6 (16\%), down to very deep limits $z'
\approx 28$, while a further two lie in the optical magnitude range
$25.5 < z' < 28$ (i.e., between the limiting magnitudes of our data
and the ACS data). The relatively low identification fraction of
Afonso et al.\ is not readily explicable considering that it is no
higher than for optical surveys which are 3-4 magnitudes
shallower. The distributions of optical magnitudes of our
identifications are shown in Fig.~\ref{fig:maghists}.
Fig.~\ref{fig:optrad} plots the $i'$ magnitude against radio flux
density, and it can be seen that there is no strong correlation
between these two variables.

\begin{table*}
\caption[]{The 100\,$\umu$Jy catalogue (only the first 20 sources are
shown). The full catalogue, containing brief notes for many sources
plus additional information, is available via the world wide web at
http://www.astro.livjm.ac.uk/\~{}cjs/SXDS/radio/, or from the authors
on request. Where a number is not listed in the Reliability column, A,
B, and C represent \textit{reliable\/}, \textit{probable\/}, and
\textit{plausible\/}, respectively, while a dash indicates a
\textit{certain\/} identification. The list is sorted in order of
decreasing flux density, and the quoted uncertainties are those
arising from the limited signal-to-noise ratio of the map, ignoring
calibration errors (we note from Fig.~\ref{fig:nvss} that our flux
density scale agrees well with that of the NVSS for the brightest
sources, where such calibration errors will be most
important).\label{tab:cat}}
\resizebox{!}{590pt}{ \rotatebox{90}{
\begin{tabular}{ccc@{~~}ccrcc@{~~}crc@{~~}c@{~~}c@{~~}c@{~~}c}
No. & Name & \multicolumn{2}{c}{Radio position} & $S_{\rm1.4\,GHz}$
& Reliability & Catalogue ID & \multicolumn{2}{c}{Optical position}
& $z_{\rm spec}$ & $B$ & $V$ & $R$ & $i'$ & $z'$ \\
& & \multicolumn{2}{c}{(J2000.0)} & ($\umu$Jy) & & &
\multicolumn{2}{c}{(J2000.0)} \\
\hline
0001 & VLA~J021827$-$04546 & 2:18:27.32 & $-$4:54:37.29 &
80250.0$\pm$72.8 & A & RC$-$124990 & 02:18:27.161 & $-$04:54:41.60 & &
23.1069 & 21.9154 & 20.9715 & 20.2001 & 19.7440 \\
0002 & VLA~J021817$-$04461 & 2:18:18.16 & $-$4:46:07.23 &
62110.0$\pm$63.9 & 0.9998 & zN$-$018551 & 02:18:18.139 &
$-$04:46:07.55 & & 26.0638 & 25.2152 & 24.3091 & 23.3194 & 22.4120 \\
0003 & VLA~J021839$-$04418 & 2:18:39.53 & $-$4:41:50.10 &
50820.0$\pm$70.9 & A & RN$-$045293 & 02:18:39.559 & $-$04:41:49.57 & &
23.2065 & 23.2808 & 23.4545 & 23.2485 & 22.9986 \\
0004 & VLA~J021853$-$04475 & 2:18:53.64 & $-$4:47:35.09 &
16950.0$\pm$67.6 & 0.2545 & RC$-$181189 & 02:18:53.587 &
$-$04:47:36.29 & & 24.9128 & 24.0525 & 24.4167 & 24.1701 & 23.5582 \\
0005 & VLA~J021851$-$05090 & 2:18:51.34 & $-$5:09:00.60 &
16010.0$\pm$65.9 & 0.9816 & RC$-$034792 & 02:18:51.326 &
$-$05:09:01.53 & & 24.5141 & 23.9246 & 23.9683 & 23.4888 & 22.9546 \\
0006 & VLA~J021637$-$05154 & 2:16:37.86 & $-$5:15:28.15 &
12020.0$\pm$77.1 & 0.9993 & zW$-$000000 & 02:16:37.825 &
$-$05:15:28.28 & & 24.2747 & 23.8249 & 23.4450 & 23.2288 & 22.7383 \\
0007 & VLA~J021659$-$04493 & 2:16:59.02 & $-$4:49:20.53 &
9600.0$\pm$135.0 & A & RC$-$160974 & 02:16:59.064 & $-$04:49:20.85 &
1.324 & 25.0818 & 24.6208 & 23.9604 & 22.9805 & 22.0174 \\
0008 & VLA~J021823$-$04530 & 2:18:23.99 & $-$4:53:04.10 &
9150.0$\pm$57.3 & 0.9958 & RC$-$132441 & 02:18:24.005 & $-$04:53:05.23
& 0.341? & 20.5376 & 19.8146 & 19.3770 & 19.1677 & 18.9268 \\
0009 & VLA~J021803$-$05384 & 2:18:03.37 & $-$5:38:25.00 &
8910.0$\pm$91.5 & 0.9938 & RS$-$004164 & 02:18:03.420 & $-$05:38:25.43
& & 26.1566 & 24.5809 & 24.5523 & 24.2270 & 24.6911 \\
0010 & VLA~J021850$-$04585 & 2:18:50.55 & $-$4:58:32.00 &
8120.0$\pm$61.3 & 0.9322 & RC$-$103643 & 02:18:50.491 & $-$04:58:32.31
& & 26.4387 & 24.5014 & 25.8306 & 25.0711 & 24.1989 \\
0011 & VLA~J021823$-$05250 & 2:18:23.52 & $-$5:25:00.44 &
7950.0$\pm$97.2 & 1.0000 & RS$-$087802 & 02:18:23.532 & $-$05:25:00.69
& 0.644 & 22.5712 & 21.3233 & 20.4645 & 19.5930 & 19.1252 \\
0012 & VLA~J021634$-$04550 & 2:16:34.99 & $-$4:55:05.61 &
6590.0$\pm$156.0 & 0.9986 & RW$-$058187 & 02:16:34.968 &
$-$04:55:06.47 & & 24.8528 & 23.4401 & 22.2077 & 21.1867 & 20.4675 \\
0013 & VLA~J021616$-$05128 & 2:16:16.82 & $-$5:12:53.47 &
6250.0$\pm$89.6 & 0.9999 & RW$-$086319 & 02:16:16.822 & $-$05:12:53.71
& 2.710 & 23.6996 & 23.0319 & 23.1279 & 23.0045 & 22.6839 \\
0014 & VLA~J021752$-$05053 & 2:17:52.135 & $-$5:05:21.25 &
6190.0$\pm$45.6 & 0.9653 & RC$-$059317 & 02:17:52.118 & $-$05:05:22.23
& & 30.1862 & 27.2512 & 26.8674 & 26.5090 & 25.8446 \\
0015 & VLA~J021932$-$05075 & 2:19:32.20 & $-$5:07:32.66 &
6100.0$\pm$117.5 & A & RE$-$068363 & 02:19:32.201 & $-$05:07:32.67 &
0.344 & 20.5793 & 19.4431 & 18.9311 & 18.7136 & 18.4210 \\
0016 & VLA~J021826$-$04597 & 2:18:26.15 & $-$4:59:46.25 &
5950.0$\pm$83.8 & B & RC$-$095238 & 02:18:25.958 & $-$04:59:45.58 &
1.132 & 24.2799 & 23.8575 & 23.8548 & 23.4662 & 22.9034 \\
 & & & & & C & RC$-$095351 & 02:18:26.088 & $-$04:59:46.76 & & 25.3795
& 24.6803 & 24.3634 & 23.8498 & 23.4315 \\
0017 & VLA~J021827$-$05348 & 2:18:27.57 & $-$5:34:53.77 &
4970.0$\pm$81.3 & $-$ & RS$-$026242 & 02:18:27.312 & $-$05:34:57.40 &
2.579 & 22.7569 & 22.4935 & 22.4985 & 22.5523 & 22.0505 \\
0018 & VLA~J021724$-$05128 & 2:17:24.38 & $-$5:12:51.68 &
4840.0$\pm$146.3 & 0.9946 & RS$-$174900 & 02:17:24.415 &
$-$05:12:52.63 & 0.918 & 23.9318 & 22.8145 & 22.1173 & 21.1717 & 20.2797 \\
0019 & VLA~J021757$-$05279 & 2:17:57.26 & $-$5:27:55.82 &
4830.0$\pm$115.5 & B & RS$-$070109 & 02:17:57.288 & $-$05:27:55.88 &
0.694 & 21.9134 & 21.7338 & 22.4154 & 21.6429 & 21.0130 \\
0020 & VLA~J021800$-$04499 & 2:18:00.68 & $-$4:49:54.78 &
4600.0$\pm$83.0 & C & zC$-$000000 & 02:18:00.716 & $-$04:49:56.41 & &
27.4331 & 26.6552 & 26.6517 & 25.8663 & 25.0541 \\
\hline
\end{tabular}
}}
\end{table*}

\begin{figure*}
\resizebox{\hsize}{!}{\includegraphics{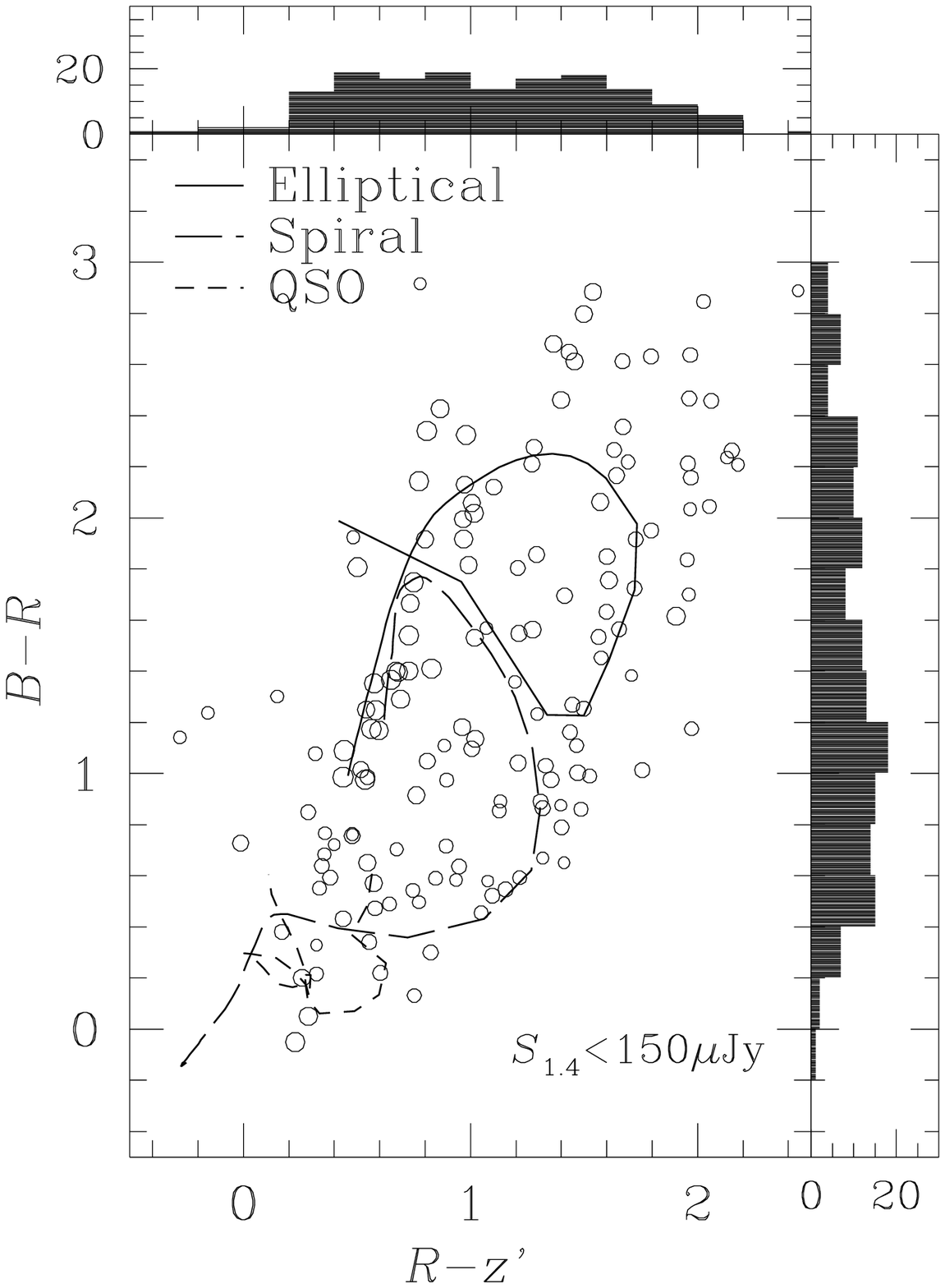}
\includegraphics{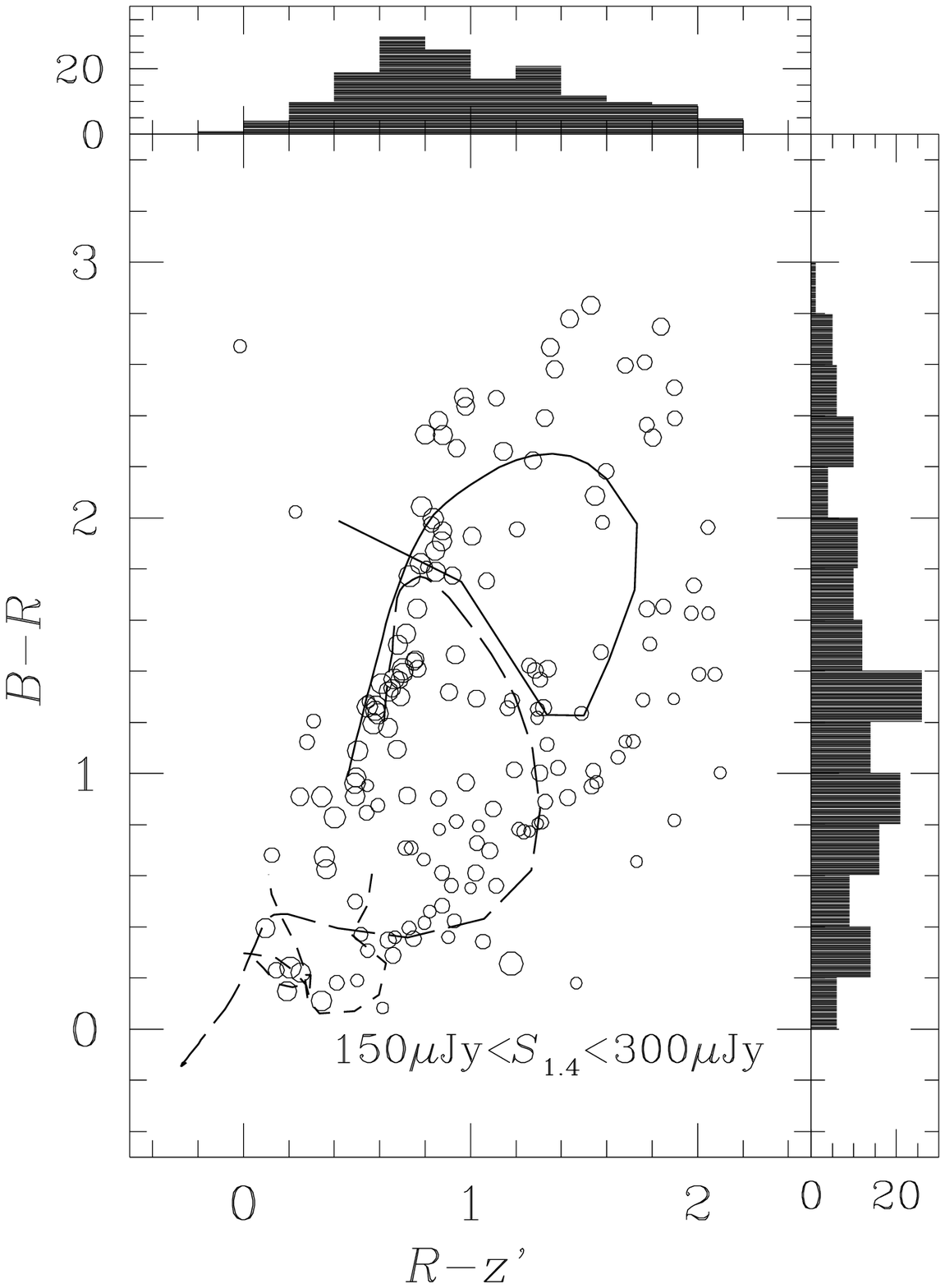}
\includegraphics{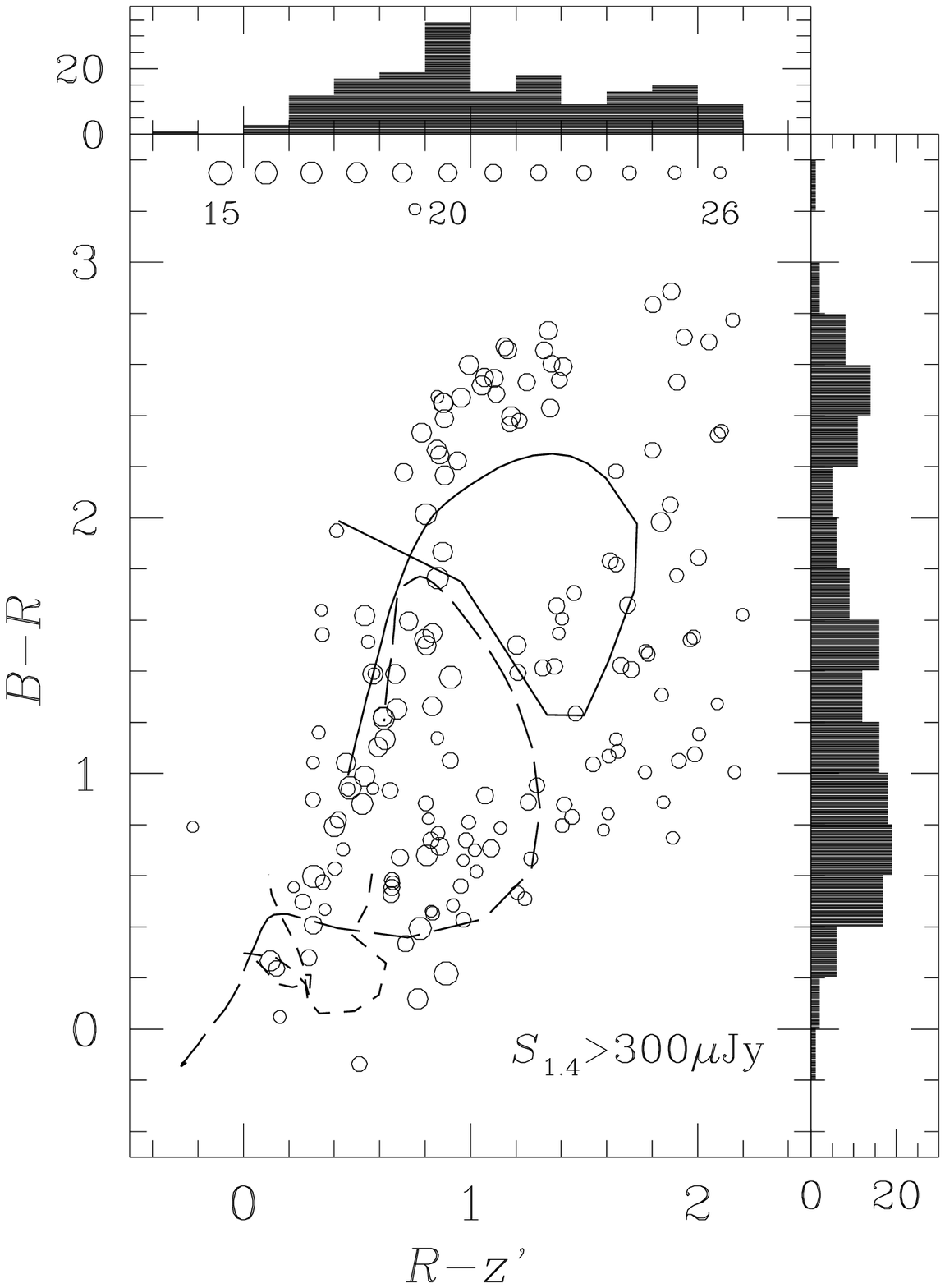}}
\caption[]{Colour--colour plots for the radio source
IDs, in different radio flux density ranges. The sizes of the symbols
represent the optical ($R$-band) magnitudes of the sources (larger
symbols represent brighter sources as shown in the right-hand panel),
and the three curves show the tracks for elliptical and spiral
galaxies, and a (non-evolving) QSO in the range $0<z<3$. The galaxy
tracks were computed using Version~2.0 of the PEGASE code (Fioc \&
Rocca-Volmerange 1997). The elliptical galaxy was modelled as an
instantaneous burst at $z=5$, while the spiral galaxy formed at $z=3$
and had an exponentially-decaying star formation rate. In both cases,
a Kroupa (2001) initial mass function was adopted and extinction was
added assuming a spheroidal geometry. The QSO track was derived from
the Sloan Digital Sky Survey composite spectrum (Vanden Berk et al.\
2001). The histograms show the distribution of colour along each
axis.\label{fig:colcol}}
\end{figure*}

Colour--colour plots for our sources were constructed, in three
different bins of radio flux density. These are shown in
Fig.~\ref{fig:colcol}, and compared to evolutionary tracks for
elliptical and spiral galaxies and a (non-evolving) QSO spectrum.
Although it is clear from these figures that photometric redshifts can
be determined for many of the radio sources from their optical colours
alone, we defer discussion of this to a later paper, for three
reasons. First, a photometric redshift code is being developed
(Furusawa et al., in preparation) which will use already-obtained
spectroscopy of galaxies in the SXDF as a training set to improve its
accuracy. Second, deep near-infrared imaging data will be obtained in
the very near future as part of the UKIDSS UDS (Lawrence et al.\
2006), and it has been demonstrated that photometric redshifts are
consistently more reliable with the inclusion of such data (e.g.,
Connolly et al.\ 1997). Finally, extensive spectroscopy has been
obtained using VLT/VIMOS and we expect this to yield spectroscopic
redshifts for a large fraction of the sources; we therefore believe it
is of little benefit to derive photo-$z$'s for our sources at this
juncture.

\subsection{Nature of the faint radio source population}

As Fig.~\ref{fig:counts} clearly shows, there is a dramatic change in
the slope of the source counts at flux densities below about 1\,mJy,
suggesting the emergence of a new population of radio
sources. Historically, it has been assumed that this new population is
low-to-moderate redshift star forming galaxies, and the overall
distribution has been successfully modelled with a suitable
parametrization of the evolution of these objects (e.g., Seymour et
al.\ 2004). However, there is no strong observational evidence to
support this as there has been limited optical identification of these
sources. Jarvis \& Rawlings (2004) have made the alternative
suggestion that the new population comprises radio-quiet AGNs. Although
we have spectroscopic redshifts for only a few radio sources in the
SXDF, the extremely high identification fraction allows us to compare
the photometric properties of the radio source IDs at different radio
flux densities.

\begin{table}
\caption[]{Comparison of the colour--colour plots shown in
Fig.~\ref{fig:colcol}. Column and rows indicate the radio flux density
ranges of the two samples being compared. The numbers are the
probabilities that the null hypothesis (that the two samples have the
same distribution) is correct, using the two-dimensional
Kolmogorov--Smirnov test of Peacock (1983), with the upper-right being
for \textit{BRz$'$\/} colours and the lower-left for
\textit{Bi$'$z$'$\/}. The parenthesized values indicate where the
probability estimation is unreliable.\label{tab:2dks}}
\centering
\begin{tabular}{rccc}
& 100--150 & 150--300 & $>$300 \\ \hline
100--150 & $\ldots$ & 0.144 & (0.791) \\
150--300 & 0.113 & $\ldots$ & 0.197 \\
$>$300 & (0.621) & 0.102 & $\ldots$ \\
\hline
\end{tabular}
\end{table}

We use the two-dimensional Kolmogorov--Smirnov test (Peacock 1983) to
investigate differences between pairs of colour--colour distributions
shown in Fig.~\ref{fig:colcol}, and summarize our results in
Table~\ref{tab:2dks}. It is clear that there is no significant change
in the properties of the optical IDs with radio flux density. We
stress that this result is robust and not dependent on the choice of
bins; for example, there is no statistically significant difference
(even at 80\,per cent confidence) between the properties of radio
sources with $100\,\umu{\rm Jy} \leq S_{1.4} < 120\,\umu\rm Jy$ and
$S_{1.4} > 1\rm\,mJy$. All three bins show a significant number of
both blue and red objects, although the central flux density bin has a
larger fraction of blue objects and is more dissimilar from the other
two bins than they are from each other. This argues against there
being a gradual change in the composition of the radio source
population as one moves to fainter flux densities, as would be
expected if starburst galaxies became increasingly dominant. Since we
know that the bright radio sources are predominantly
passively-evolving massive elliptical galaxies (e.g., Willott et al.\
2003), many of the faint radio sources must also be ellipticals.

\begin{figure}
\resizebox{\hsize}{!}{\includegraphics{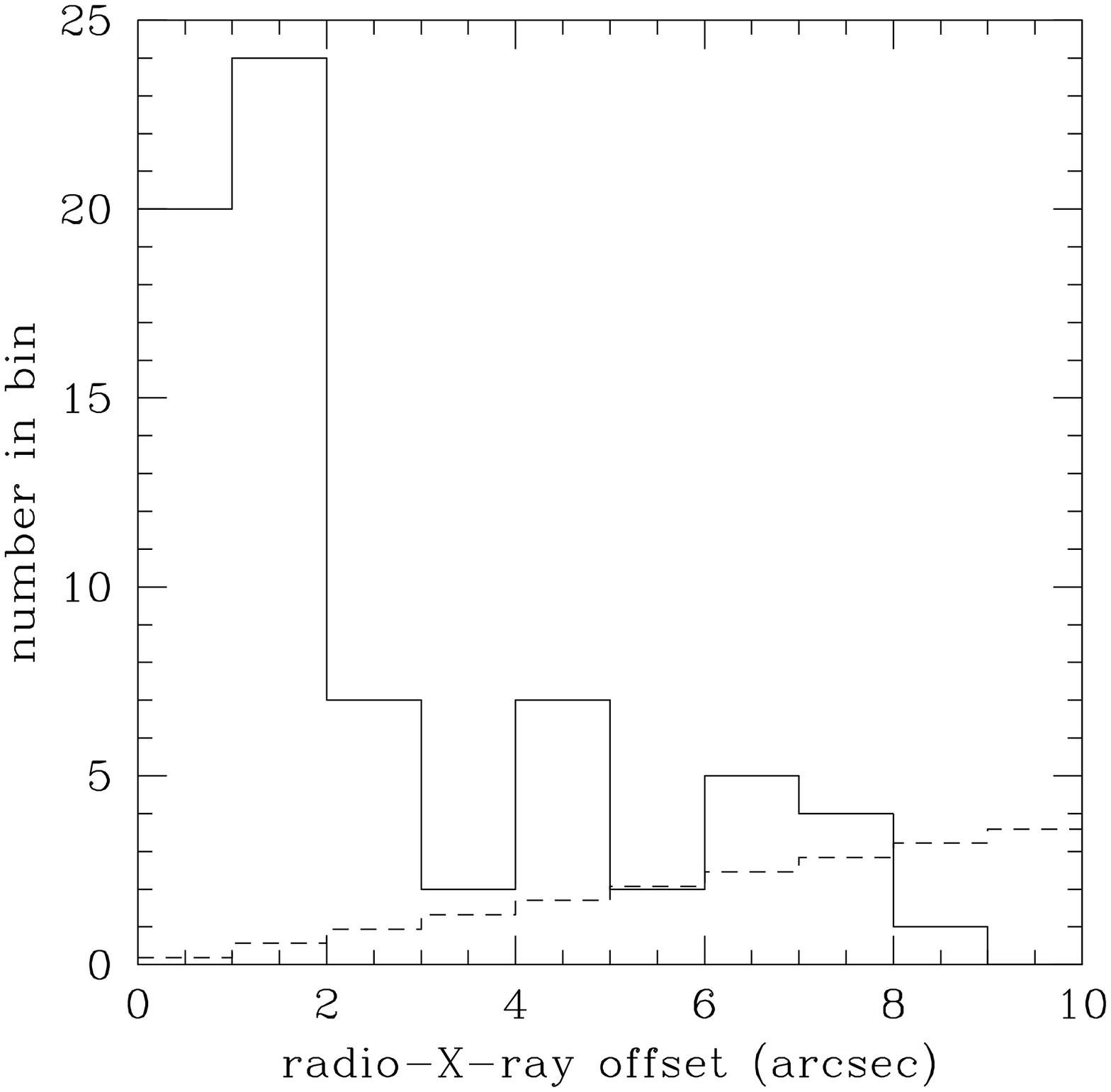}}
\caption[]{Histogram of the separations between the locations of the optical
counterparts of radio sources and the X-ray sources to which they have been
matched. The dashed line shows the expectation of chance associations
assuming the X-ray sources are distributed uniformly on the sky. We
consider sources to be associated if the separation is
$<5''$.\label{fig:xradoff}}
\end{figure}

We investigate further by comparing the radio and X-ray properties of
our radio sources. We use the X-ray catalogue of Ueda et al.\ (2006)
and apply the same detection thresholds as Akiyama et al.\ (2006),
namely that a source must have a likelihood ${\cal L} \geq 9$ in one
<of the 0.3--10\,keV (total), 0.5--2\,keV (soft), or 2--10\,keV (hard)
bands. These are fairly conservative criteria in the sense that only 3
spurious sources are expected in the total band sample per
\textit{XMM-Newton\/} pointing. We search for the X-ray source nearest
to the position of each optical counterpart (or the radio position if
there is no optical counterpart), and find the distribution of
optical--X-ray separations shown in Fig.~\ref{fig:xradoff}. From this
Figure, we decide to consider radio and X-ray sources to be
counterparts if the separation is less than 5\,arcseconds, which
results in 60 radio--X-ray matches (i.e., $11.9\pm1.4$\% of the radio
source population), of which 6 are expected to be chance
associations. None of our X-ray matches has a significant spatial
extent.

Relationships have been determined between the star formation rate (of
stars with $M>5M_\odot$) and the radio and X-ray luminosities of
star-forming galaxies: (Condon 1992; Ranalli, Comastri \& Setti 2003),
which can be combined to determine the relationships between radio and
X-ray luminosities for star-forming galaxies. Since the spectral
shapes in the radio and X-ray regimes are very similar ($S_\nu \propto
\nu^{-0.8}$), these can be used to derive a correlation between fluxes
which can be applied to our data where redshifts (and hence
luminosities) are unknown. The relationships are:
\begin{eqnarray}
S_{\rm0.5-2keV} ({\rm W\,m^{-2}}) & = & 10^{-18.0} S_{\rm1.4GHz} ({\rm
mJy})
\\
S_{\rm2-10keV} ({\rm W\,m^{-2}}) & = & 10^{-18.0} S_{\rm1.4GHz} ({\rm mJy})
\end{eqnarray}

Correlations between X-ray and radio luminosities for radio-quiet AGNs
have been determined by Brinkmann et al.\ (2000). These are very close
to being proportionalities, and so we can convert them to
relationships between fluxes:
\begin{eqnarray}
S_{\rm0.5-2keV} ({\rm W\,m^{-2}}) & = & 10^{-15.5} S_{\rm1.4GHz} ({\rm
mJy})
\\
S_{\rm2-10keV} ({\rm W\,m^{-2}}) & = & 10^{-15.3} S_{\rm1.4GHz} ({\rm mJy})
\end{eqnarray}

We also consider the correlation which is present between the radio
core and X-ray luminosities for low-power (FR\,I) radio galaxies
(Canosa et al.\ 1999; FR\,II radio sources would need to lie at $z>4$
to be this faint). Hardcastle \& Worrall (2000) have attributed this
to an inverse Compton origin for the X-ray emission. If we assume that
our measured radio flux densities are dominated by the flat spectrum
core components of FR\,I radio galaxies, we expect them to obey the
following correlations:
\begin{eqnarray}
S_{\rm0.5-2keV} ({\rm W\,m^{-2}}) & = & 10^{-19.3} (1+z)^{-0.8}
S_{\rm1.4GHz} ({\rm mJy})
\\
S_{\rm2-10keV} ({\rm W\,m^{-2}}) & = & 10^{-19.1} (1+z)^{-0.8}
S_{\rm1.4GHz} ({\rm mJy})
\end{eqnarray}
Only the brightest, most nearby sources would be detected in the SXDF
X-ray images, which have a flux limit of $\sim
10^{-18}\rm\,W\,m^{-2}$, and we do not consider this mechanism further.

\begin{figure*}
\resizebox{\hsize}{!}{\includegraphics{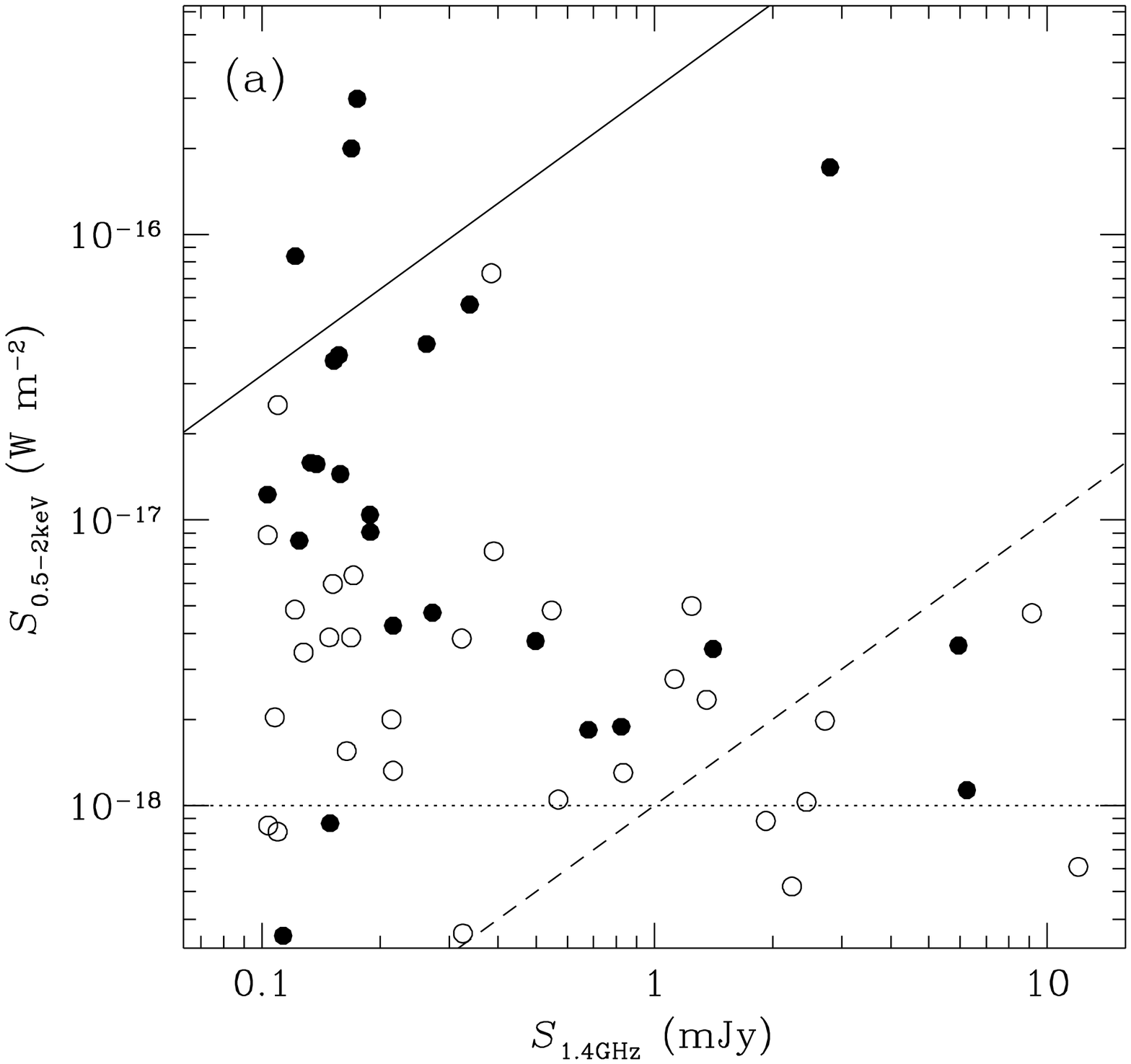}
\includegraphics{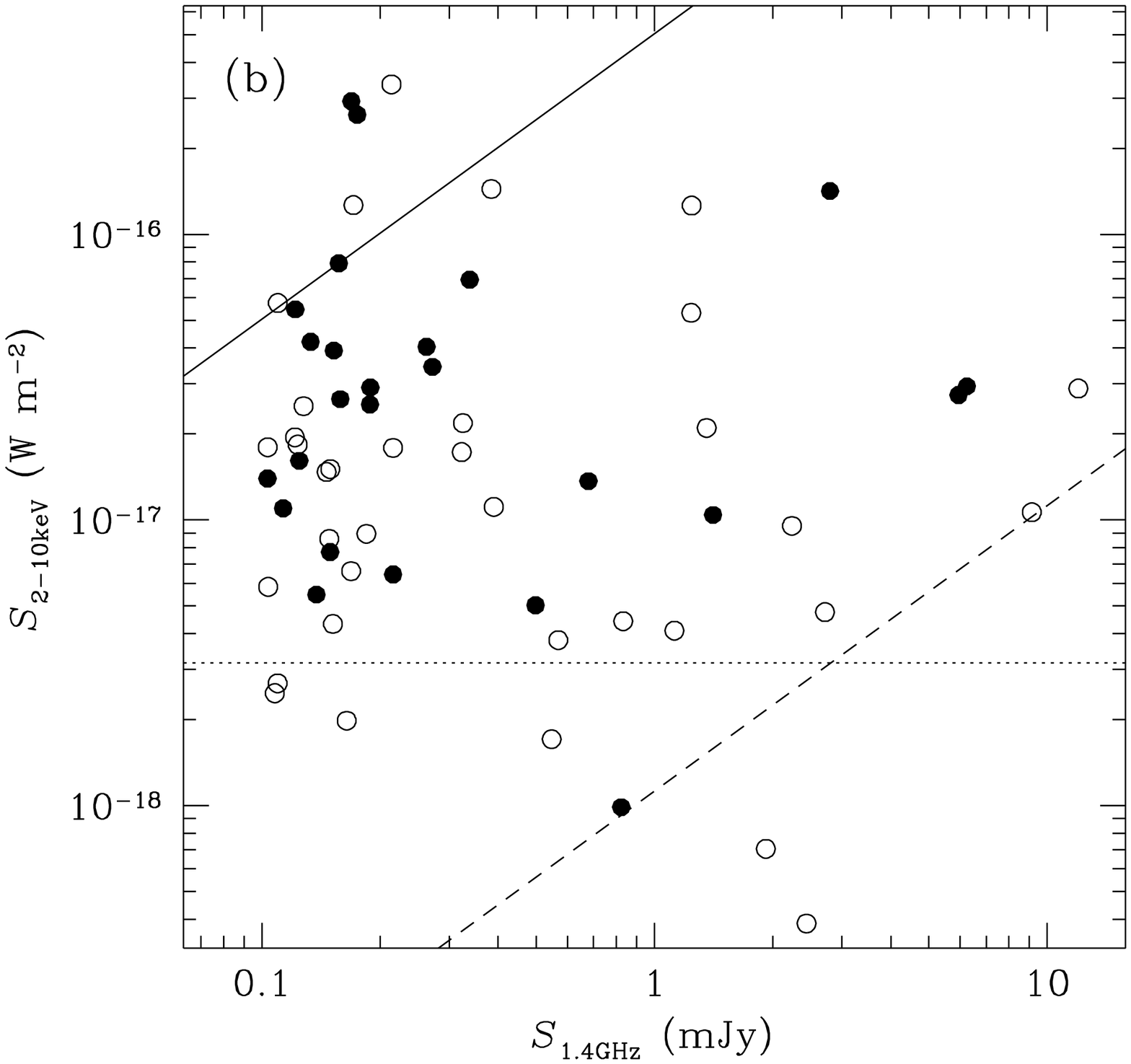}}
\caption[]{Plots of (a) soft and (b) hard X-ray flux vs radio flux
density for the radio sources with X-ray detections. Filled symbols
represent optical point sources and open symbols optically extended
objects. In each plot, the dashed line indicates the relationship for
starburst galaxies (Eqns~6--7) while the solid line shows the
correlation for radio-quiet AGNs (Eqns~8--9). Most (88\,per cent) of
the radio sources are not detected in X-rays; we show the approximate
limits of the X-ray data (Akiyama et al.\ 2006) by the dotted lines in
each figure.\label{fig:xradio}}
\end{figure*}

We plot the soft and hard X-ray fluxes (computed assuming an X-ray
photon index $\Gamma=1.8$) of the 60 X-ray-detected radio sources
versus radio flux density in Fig.~\ref{fig:xradio}. It is immediately
clear from these plots that a population of X-ray bright radio sources
appears at radio fluxes $S_{1.4} \la 300\,\umu\rm Jy$, i.e., where the
new population of radio sources starts to dominate. The ratios of
X-ray flux to radio flux density of these sources indicate that they
are radio-quiet AGNs.

The optical morphologies of these objects provide further clues as to
their nature. We use the SExtractor {\tt CLASS\_STAR} field (Bertin \&
Arnouts 1996) to determine whether the objects are point-like or
resolved in the $i'$-band, using a value of 0.8 as the
borderline. Fig.~\ref{fig:xradio}(a) shows that most objects which are
bright in the soft X-ray band are point-like, and so are almost
certainly radio-quiet QSOs which suffer little obscuration at either
optical or X-ray wavelengths. In Fig.~\ref{fig:xradio}(b), however,
many of the hard-X-ray-bright objects are resolved optically and are
therefore likely to be `Type 2' objects, where the nucleus suffers
significant obscuration.

If we assume that objects in Fig.~\ref{fig:xradio}(b) are radio-quiet
AGNs or starbursts according to which of the two correlations they are
closer to, we can determine that at least 10\,per cent of radio
sources with $S_{1.4}<300\rm\,\umu Jy$ are radio-quiet AGNs. This
fraction is the same in the bins of radio flux density
100--150\,$\umu$Jy and 150--300\,$\umu$Jy, and is a lower limit since
there are two effects which can cause objects to migrate towards the
bottom-right of this plot: relativistic effects can boost the radio
flux of a source whose jets are oriented close to the line of sight,
and photoelectric absorption can significantly reduce the observed
X-ray flux. With regard to the first point, Falcke, Sherwood \&
Patnaik (1996) suggest that $\sim10$\,per cent of radio-quiet QSOs may
have their core radio fluxes (and hence their overall radio fluxes)
enhanced by Doppler boosting. Regarding the second point, the
2--10\,keV count rate from an AGN at $z=0.7$ is reduced by an order of
magnitude for an absorbing column $N_{\rm H} =
1.5\times10^{24}\rm\,cm^{-2}$, and the fraction of such Compton-thick
AGNs has been estimated to be 20--30\,per cent of the total, both from
observations of local Seyfert galaxies (Risaliti, Maiolino \& Salvati
1999), and from fits to the Cosmic X-ray Background (Ueda et al.\
2003). Furthermore, the correlations between radio and X-ray
luminosity used above were derived from cross-correlation of a deep
radio catalogue and a shallower X-ray catalogue, and will therefore be
biased towards unusually X-ray-luminous sources. The true correlations
for radio-quiet AGNs could be significantly lower in
Fig.~\ref{fig:xradio} and the fraction of sources which lie close to
it may have been underestimated as a result. Indeed, this seems very
likely since the solid lines in Fig.~\ref{fig:xradio} clearly exceed
the median X-ray fluxes of the QSOs. Our ongoing follow-up of both
radio- and X-ray-selected samples in the SXDF will enable an
investigation of the correlations between X-ray and radio luminosities
without such biases. We tentatively suggest that 20\,per cent or more
of the radio sources with $100\,\umu{\rm Jy} \leq S_{1.4} <
300\,\umu\rm Jy$ are radio-quiet AGNs. Models for the evolution of the
radio luminosity function indicate that, despite their faintness,
approximately half of the radio sources at these flux densities are
radio-loud AGNs (e.g., Jarvis \& Rawlings 2004; this conclusion rests
on the assumption that there is not a strong cut off in the space
density of low-power FR\,I radio sources at $z\la1$), and therefore
radio-quiet AGNs may provide a larger fraction of the flattening
signal than starburst galaxies.

This has clear implications for studies of the Cosmic X-ray Background
(CXB), as alluded to by Jarvis \& Rawlings (2004). These authors
considered the reverse calculation, using the hard X-ray luminosity
function of Ueda et al.\ (2003) to predict the contribution from
radio-quiet AGNs to the radio source counts at faint flux
densities. From their fig.~3, $\sim10$--15\,per cent of radio sources
with $100\,\umu{\rm Jy} \leq S_{1.4} < 300\,\umu\rm Jy$ are
radio-quiet AGNs, which is in agreement with our lower limit. However,
if there is a population of Compton-thick AGNs, these will not be seen
in the 2--10\,keV source counts of Ueda et al., and hence Jarvis \&
Rawlings have also calculated a lower limit to the radio-quiet AGN
fraction at faint radio flux densities. While such Compton-thick
objects do not contribute much of the 2--10\,keV background light,
they will not be as strongly absorbed at the $\sim30$\,keV peak of the
CXB. Ueda et al.'s (2003) model of the CXB has a knee in the 30\,keV
source counts where most of the background is produced, and these
objects have intrinsic 2--10\,keV fluxes near the observed knee in the
2--10\,keV counts at $S_{\rm2-10keV} \approx 2 \times
10^{-17}\rm\,W\,m^{-2}$, or $S_{1.4} \approx 40\,\umu\rm Jy$. While
the observed X-ray flux may be reduced by a large (possibly
Compton-thick) absorption column, the radio flux will remain
unaffected, and hence radio surveys to this depth can identify the
objects which dominate at the 30\,keV peak of the CXB, \textit{even if
they are Compton-thick and invisible to deep X-ray surveys\/}. While
this radio-quiet AGN population will be contaminated by star-forming
galaxies, it can be expected that colour-selection criteria will allow
discrimiation between these two populations (e.g., by using the
\textit{Spitzer\/}/IRAC bands to detect the AGN-heated dust at
$T\sim1000$\,K which is not present in starburst galaxies), and hence
separate the much brighter (at X-ray energies) AGNs from the
starbursts. One can hence determine unambiguously whether a population
of intrinsically luminous obscured high-redshift radio-quiet AGNs does
exist, even if they are too faint for current X-ray telescopes. This
is in marked contrast to the results from deep X-ray surveys, where
the increased depth has resulted in the detection of low-redshift,
apparently intrinsically faint, AGNs (e.g., fig.~5 of Barger et al.\
2003), rather than the hoped-for higher redshift population. It is
possible that some fraction of the faintest X-ray sources are
Compton-thick objects where only scattered radiation is being seen at
2--10\,keV, but there is no definitive way to test this except to
measure the X-ray fluxes at higher energies. We will study the
multiwavelength properties of the faint radio sources in a later
paper.

\section{Summary}

We have presented a radio map covering 1.3 square degrees in the region of
the Subaru/\textit{XMM-Newton\/} Deep Field, and described the production
of a catalogue covering 0.81 square degrees containing all sources with a
peak flux density greater than 100\,$\umu$Jy\,beam$^{-1}$. We have made
reliable identifications for 90\,per cent of the sources, which is a
significantly higher fraction than other surveys covering a similar area.
We have demonstrated that there is no significant difference between the
optical colours of radio sources selected at different flux densities, and
have combined this with X-ray data to show that an important fraction of
the faint ($100\,\umu{\rm Jy} \leq S_{1.4} < 300\,\umu\rm Jy$) radio source
population must be radio-quiet AGNs, including Type~2 objects which
contribute to the Cosmic X-ray Background.

\section*{Acknowledgments}

This paper is partially based on data collected at Subaru Telescope,
which is operated by the National Astronomical Observatory of
Japan. The National Radio Astronomy Observatory is a facility of the
National Science Foundation operated under cooperative agreement by
Associated Universities, Inc. CS and SR thank the Particle Physics and
Astronomy Research Council for funding in the form of an Advanced
Fellowship and Senior Research Fellowship, respectively. The authors
thank Nick Seymour for providing the data used to produce
Fig.~\ref{fig:counts}, and the referee, Jim Condon, for numerous
helpful comments.

\appendix

\section{Extended sources}

\begin{figure*}
\includegraphics[scale=1.15]{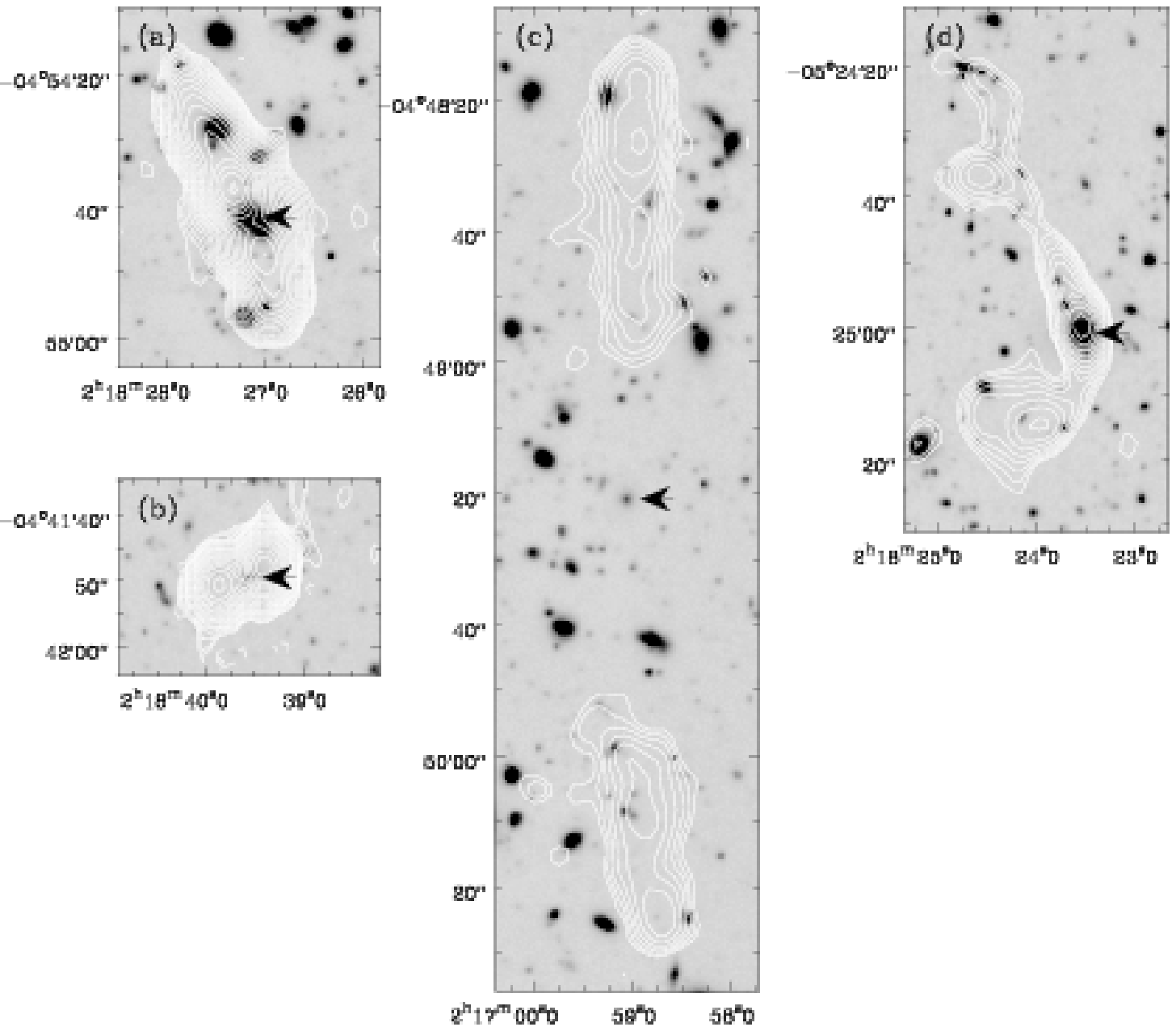}
\caption[]{Radio/optical overlays of extended radio sources in the
SXDF. Radio contours start at 50\,$\umu$Jy\,beam$^{-1}$ and are spaced at
intervals of $\sqrt{2}$. The greyscale is the Subaru/Suprime-Cam $i'$-band
image, and the proposed identification in each case is marked with an
arrow. All images are presented at the same scale. (a) VLA~0001
(J021827$-$04546) (b) VLA~0003 (J021839$-$04418) (c) VLA~0007
(J021659$-$04493) (d) VLA~0011 (J021823$-$05250).\label{fig:extended}}
\end{figure*}

\addtocounter{figure}{-1}
\begin{figure*}
\includegraphics[scale=1.15]{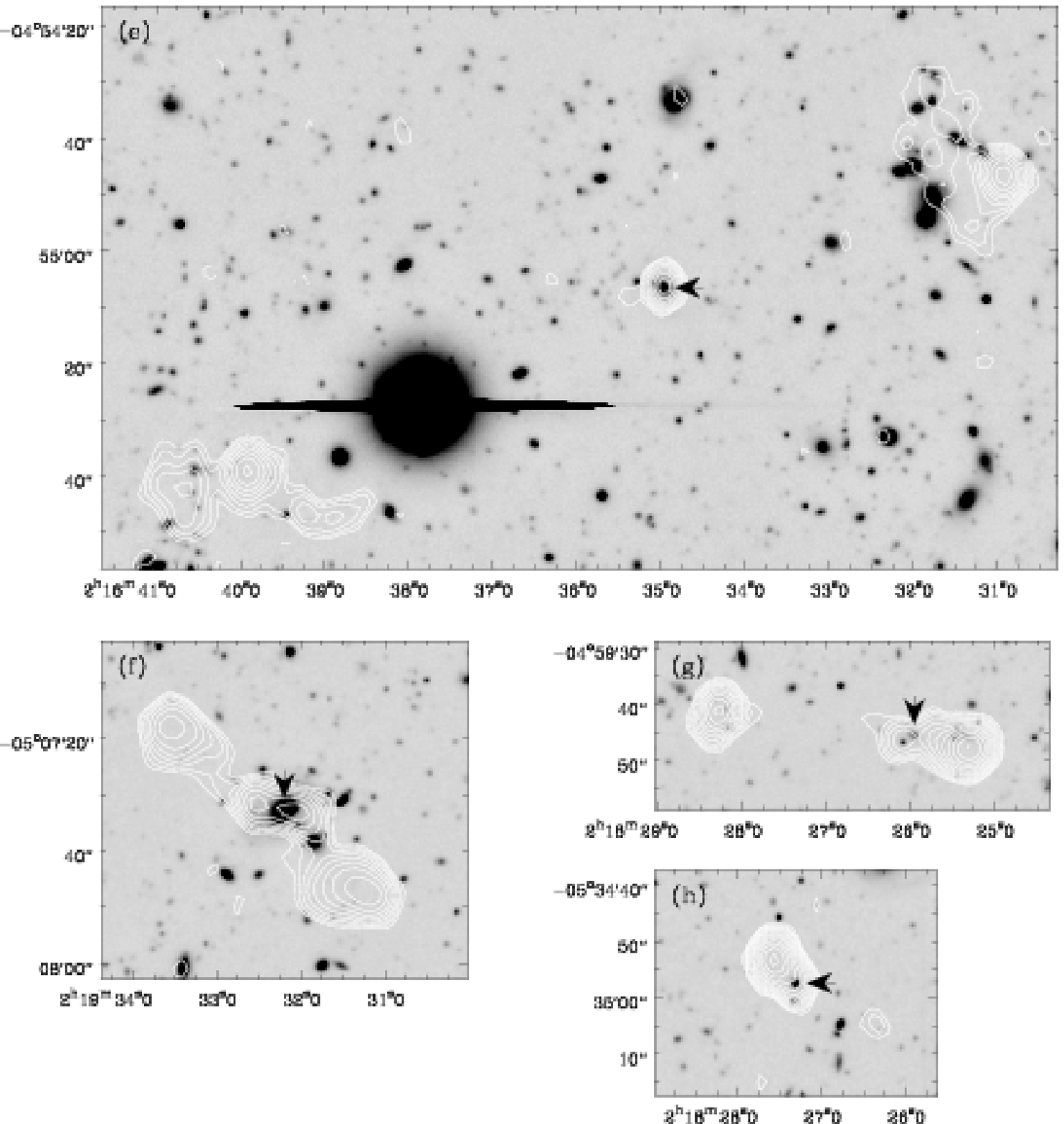}
\caption[]{\textit{continued.\/} (e) VLA~0012 (J021634$-$04550) (f)
VLA~0015 (J021932$-$05075) (g) VLA~0016 (J021826$-$04597) (h) VLA~0017
(J021827$-$050348).}
\end{figure*}

\addtocounter{figure}{-1}
\begin{figure*}
\includegraphics[scale=1.15]{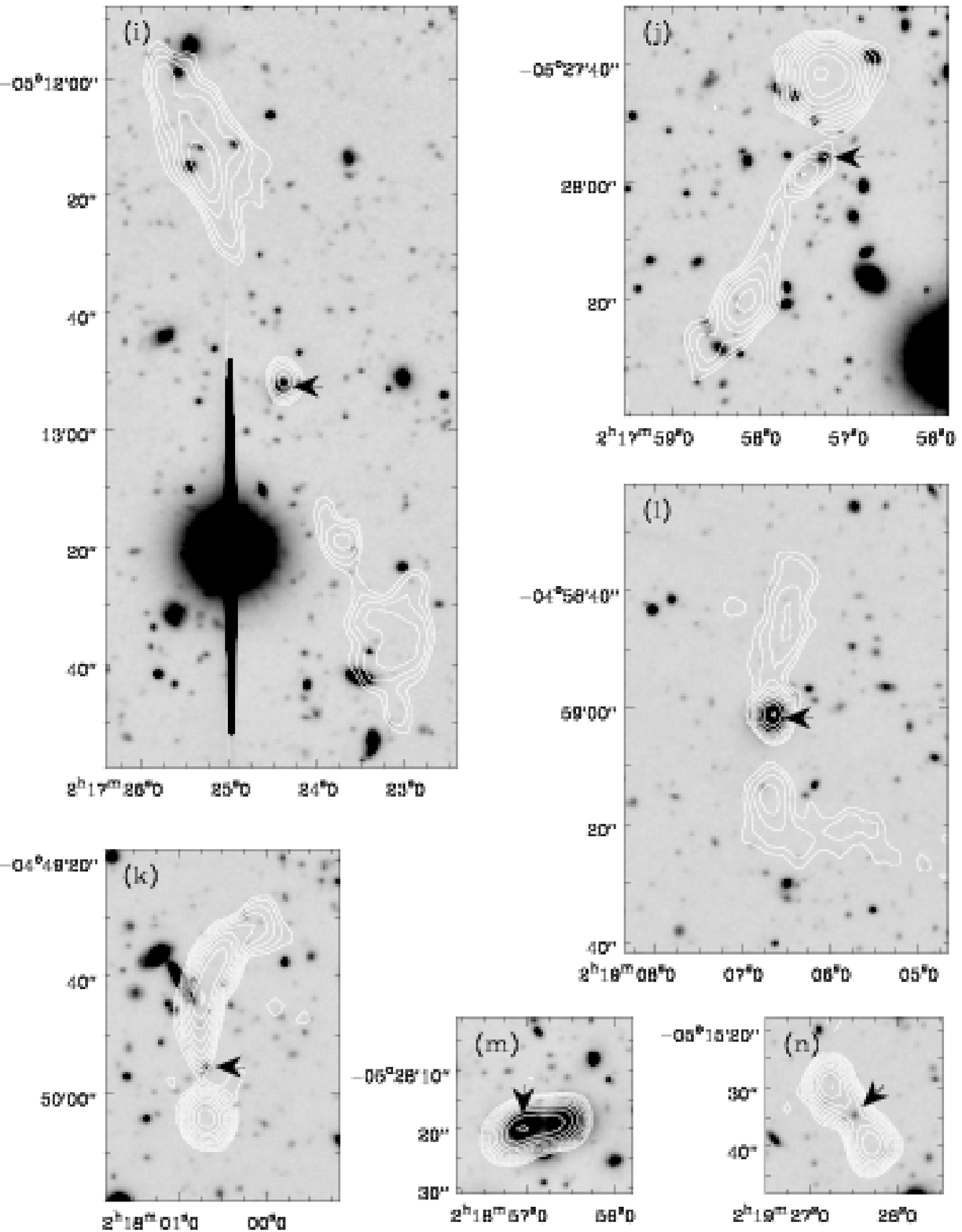}
\caption[]{\textit{continued.\/} (i) VLA~0018 (J021724$-$05128) (j)
VLA~0019 (J021757$-$05279) (k) VLA~0020 (J021800$-$04499) (l) VLA~0024
(J021906$-$04590) (m) VLA~0026 (J021856$-$05283) (n) VLA~0032
(J021926$-$05155).}
\end{figure*}

\addtocounter{figure}{-1}
\begin{figure*}
\includegraphics[scale=1.15]{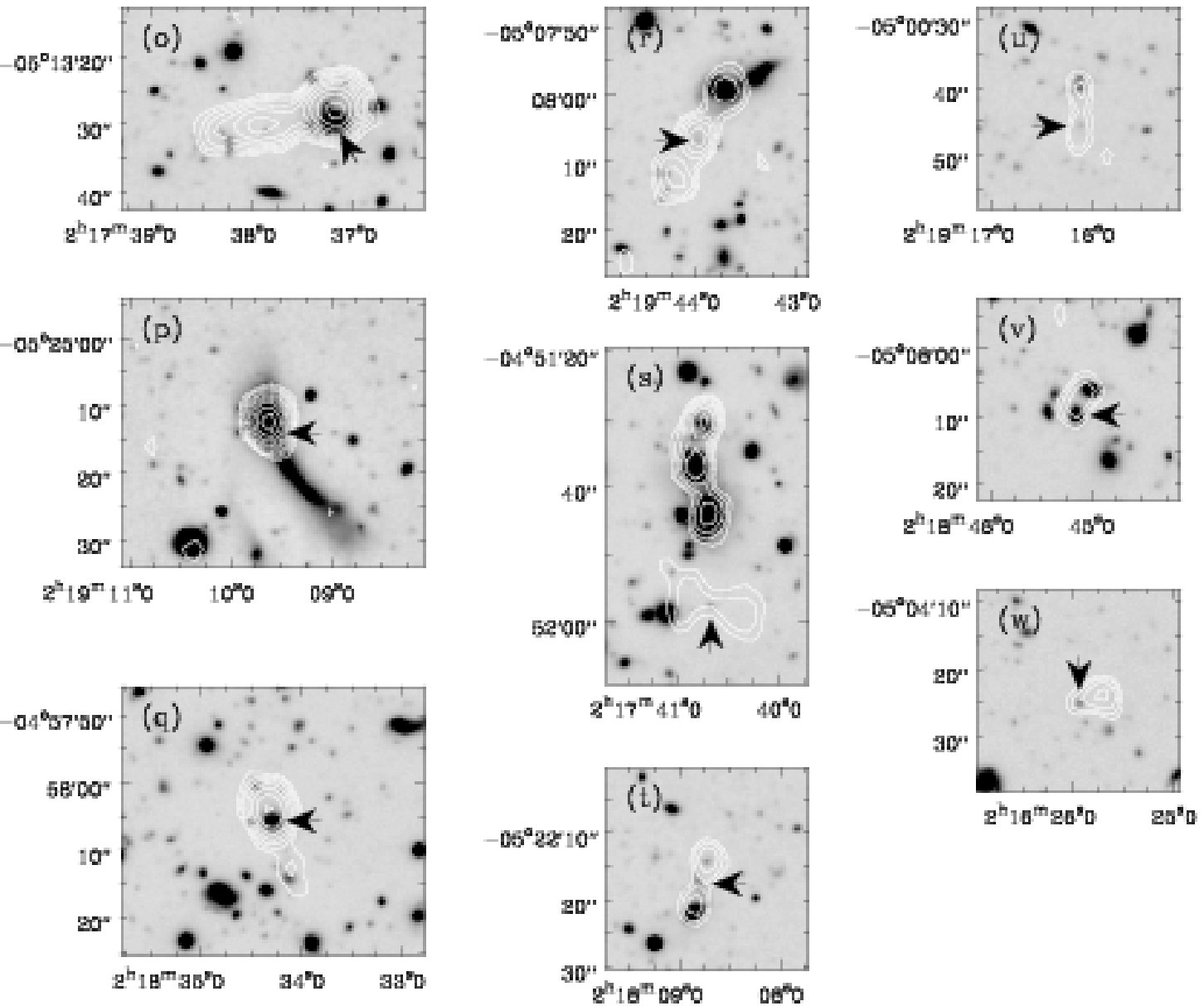}
\caption[]{\textit{continued.\/} (o) VLA~0033 (J021737$-$05134) 
(p) VLA~0049 (J021909$-$05252) (q) VLA~0067 (J021834$-$04580) (r)
VLA~0086 (J021944$-$05081) (s) VLA~0103 (J021740$-$04519), VLA~0167
(J021740$-$04515), VLA~0177 (J021740$-$04516), and VLA~0260
(J021740$-$04517) (t) VLA~0175 (J021808$-$05222) (u) VLA~0244
(J021916$-$05007) (v) VLA~0247 (J021845$-$05081) (w) VLA~0278
(J021625$-$05044).}
\end{figure*}

In this Appendix, we present radio/optical overlays of radio sources
with significantly extended morphologies (Fig.~\ref{fig:extended}),
together with our assessments of the proposed optical counterparts
(see Section~4.1). We also provide brief notes on each source. All
spectroscopic redshifts quoted are from M.~Akiyama (private
communication) unless otherwise noted. Colours are measured in
2-arcsec diameter apertures (Furusawa et al.\ 2006).

\newcounter{figgy}
\begin{list}{(\alph{figgy})}{\usecounter{figgy}}
\item VLA0001 (J021827$-$04546) is the brightest source in our
catalogue. It is an edge-darkened source \textit{reliably\/}
identified with a large elliptical galaxy.
\item VLA0003 (J021839$-$04418) is a small ($D=7.5''$) double radio source
\textit{reliably\/} identified with an extended blue object at its centre.
\item VLA0007 (J021659$-$04493) is a giant radio source
($D=129''$). Although no radio core is seen ($S_{1.4}<46\,\umu$Jy),
there is only one plausible optical ID, a very red galaxy ($R-z'=2.0$)
almost exactly midway between the radio hotspots, and we consider this
a \textit{reliable\/} identification.
\item VLA0011 (J021823$-$05250) is a wide-angle radio source, whose
bright core ($S_{1.4}=3930\pm30\,\umu$Jy) is coincident with a large
red galaxy. Such sources typically reside in galaxy clusters, and the
surrounding objects display a clear red sequence (Geach et al., in
preparation) but there is no X-ray emission (Y.~Ueda, private
communication).
\item VLA0012 (J021634$-$04550) is a large ($D=144''$) FR\,II source,
whose bright radio core ($S_{1.4}=992\pm16\,\umu$Jy) is coincident
with a red galaxy.
\item VLA0015 (J021932$-$05075) is a `double-double' radio source with a
pronounced change in position angle between the inner and outer lobes.
There is no clear core, but there is a very bright ($R\approx19$)
galaxy located between the two inner lobes, which we
\textit{reliably\/} propose as the identification.
\item VLA0016 (J021826$-$04597) is an apparently highly asymmetric (in
terms of arm length) triple. A possible optical counterpart is seen near
the central (core?) component, but an X-ray source is detected whose
position better matches the galaxy between the central and western
components. Spectroscopy of this object reveals strong [O{\sc~ii}] at
$z=1.132$, so it is considered to be a \textit{probable\/}
identification, while the object closest to the central radio
component is considered to be \textit{plausible\/}. A spectrum of this
source is desired to see if it displays the characterisitic signatures
of an active galaxy.
\item VLA0017 (J021827$-$05348) has no likely optical ID at the location of
the radio source peak, but a blue unresolved source is seen in the
southwestern extension. An optical spectrum of this reveals it to be a
$z=2.579$ QSO, and we therefore consider its identification as the
optical counterpart to be \textit{certain\/}, with the radio emission
being dominated by a Doppler-boosted jet. This interpretation is
further supported by the presence of faint radio emission on the
opposite (receding) side of the source.
\item VLA0018 (J021724$-$05128) is a giant source ($D=105''$) with a
well-detected core ($S_{1.4}=253\pm16\,\umu$Jy). It is identified with
a red galaxy at $z=0.918$.
\item VLA0019 (J021757$-$05279) has an unusual radio morphology. The only
\textit{probable\/} ID lies between the northernmost two radio peaks,
which is a red galaxy at $z=0.694$.
\item VLA0020 (J021800$-$04499) is comprised of two main radio components:
a point-like source to the south, and an extended arc-like component. There
is also evidence for faint radio emission connecting the southern source to
the northern end of the extended component. No optical object is visible at
the location of the southern radio peak, but a faint ($R>26$) red
($(R-K)_{\rm AB} \approx 3.5$); T.~Takata, private communication) object is
visible between the two radio components, and we consider this to be a
\textit{plausible\/} identification.
\item VLA0024 (J021906$-$04590) is an edge-darkened source with a
strong core ($S_{1.4}=868\pm16\,\umu$Jy) and two jets of radio
emission. It is identified with an isolated bright elliptical galaxy.
\item VLA0026 (J021856$-$05283) is associated with an elliptical
galaxy at $z=0.450$ which is the eastern member of a pair of bright
elliptical galaxies apparently at the centre of a cluster. We have
investigated the possibility of the radio flux being enhanced by
emission from the western galaxy by fitting two Gaussians to the radio
map. The fitted components have flux densities of $1.41\pm0.02$\,mJy
(eastern) and $1.22\pm0.02$\,mJy (western), but the western component
is displaced by more than 2\,arcsec from the optical location of the
galaxy, and there is a $\ll1$\,per cent probability of the western
galaxy being the correct identification for a source at this
location. We therefore conclude that this is a single radio source.
\item VLA0032 (J021926$-$05155) is a double radio source ($D=20''$)
\textit{reliably\/} associated with a red ($R-z'=2.0$) galaxy.
\item VLA0033 (J021737$-$05134) appears to be a core-jet radio
source whose proposed optical ID is a red galaxy $\sim1''$
away from the radio peak. This galaxy lies at $z=0.647$ and is part of
a cluster (Geach et al.\ 2006) and we consider it to be a
\textit{reliable\/} identification based on the radio morphology. The
source appears to be strongly lensing a blue object, but the lack of
any contaminating emission lines from this sources suggests that the
background object is unlikely to be the ID.
\item VLA0049 (J021909$-$05252) is an unresolved radio source
coincident with a tadpole-like galaxy. It is included here because the
position of the galaxy in the optical catalogue is somewhat displaced
from its nucleus (which is coincident with the radio position; see
figure) and hence the galaxy has a low formal probability of being the
correct identification. However, there can be no doubt that the
disturbed galaxy is the optical counterpart and we consider it
\textit{certain\/}.
\item VLA0067 (J021834$-$04580) is either a core-jet source, or
possibly a very asymmetric double (the northern peak has a flux
density more than 5 times that of the southern peak). If the brighter
peak is associated with the core, then the $R\approx22$ galaxy
indicated has a formal probability of 4.5\% of being the correct
identification.
\item VLA0086 (J021944$-$05081) is a triple radio source of angular
extent $8''$. The nature of this source is ambiguous since there are
optical counterparts at the locations of both the central and northern
radio peaks. However, a chance alignment of three radio sources is
improbable, we believe that all three components form a single source,
and can provide a formal probability for an identification based on
the radio position of the $R\approx23$ central source, which is
99.97\%. The $R\approx20$ galaxy coincident with the northern radio
source may contribute to its radio flux.
\item This is a montage of several distinct radio sources. The brightest
(VLA0103 = J021740$-$4519) is a coreless double \textit{reliably\/}
identified with a 25th magnitude galaxy at its centre. To the north
are three radio sources (VLA0167 = J021740-04515; VLA0177 =
J021740$-$04516; VLA0260 = J021740$-$04517), each coincident with a
relatively bright ($i'\approx19$--21) galaxy which have high formal
probabilities of being the correct identifications.
\item VLA0175 (J021808$-$05222) also cannot be uniquely identified, as
there are optical sources coincident with both radio components, as well as
midway between them. Given the very similar flux densities of the two
components, we consider it most likely to be a double radio source and
therefore propose the central galaxy as a \textit{probable\/}
identifcation.
\item VLA0244 (J021916$-$05007) appears to be a double radio source,
although there is an additional extension to the southwest of the
southern peak. The optical ID is highly uncertain, but we consider the
25th magnitude galaxy between the two main peaks to be a
\textit{plausible\/} identification.
\item VLA0247 (J021845$-$05081) is an extended radio source coincident with
two galaxies. The peak of the radio emission indicates that the
southern galaxy is the identification, but the northern galaxy may
contribute some radio flux.
\item VLA0278 (J021625$-$05044) possesses extended emission to the
east of the radio peak. A 24th magnitude object lies within the
eastern extension of the radio source, and this is considered to be a
\textit{plausible\/} identification as the position of the radio peak
may not be indicative of the location of the source.
\end{list}

\bsp

\clearpage

\label{lastpage}
\end{document}